\documentclass[aps,twocolumn]{revtex4}
\usepackage{color}
\usepackage{psfrag}
\usepackage{dcolumn}
\usepackage{bm}
\usepackage{float}
\usepackage[latin1]{inputenc}
\usepackage[spanish,english]{babel}
\usepackage{amsfonts}
\usepackage{amssymb,epsf}
\usepackage{graphicx}
\usepackage{epstopdf}
\usepackage{amsmath,amssymb}
\usepackage{pdflscape}
\usepackage{adjustbox}
\usepackage{hyperref}
\usepackage{subcaption}
\begin{document}

\title{Constraint on the equation of state of strange quark star: Perturbative QCD along with a density-dependent bag constant}

\author{J. Sedaghat\footnote{
		email address: J.sedaghat@shirazu.ac.ir}, G. H. Bordbar\footnote{
		email address: ghbordbar@shirazu.ac.ir (corresponding author)}, S. M. Zebarjad\footnote{
		email address: zebarjad@shirazu.ac.ir (corresponding author)}}

\affiliation{Physics Department and Biruni Observatory, Shiraz University, Shiraz 71454, Iran}
	
\begin{abstract}

This study investigates the structural properties of strange quark stars (SQS) using a Quantum Chromodynamics (QCD) perturbative model combined with the latest Particle Data Group dataset. Given the energy scale present in compact stars, QCD perturbation theory alone may not fully explain their structure. To account for non-perturbative contributions, we incorporate a density-dependent effective bag parameter, $B$, and derive the equation of state (EOS) for strange quark matter (SQM). We start by demonstrating the limitations of EOSs with a constant $B$ in describing massive objects with $ M_{TOV}> 2M_{\odot} $. Subsequently, we show that considering $B$ as a  density-dependent function significantly changes the results. Our definition of $B$ includes two parameters determined by both theoretical and observational constraints. We demonstrate that incorporating a density-dependent $B$ into the perturbative EOS can yield SQSs with masses exceeding $2M_{\odot}$, while complying with gravitational wave constraints such as tidal deformability, and thermodynamic considerations, including stability conditions and speed of sound behavior. Specifically, we show that massive compact objects like PSR J0952-0607, PSR J2215+5135, PSR J0740+6620, and the secondary mass of GW190814 can be SQSs. {Additionally, we compare our EOS with the EOS of the authors who use a generalized polytropic form with adjustable parameters and obtain an interesting result.}


\end{abstract}

\maketitle

\section{Introduction}
Gravitational waves have reshaped our understanding of compact objects such as neutron stars and black holes, offering valuable insights into their origins, evolution, and structure. 
The neutron star binary mergers detected by LIGO and Virgo have provided crucial information on the masses, spins, and tidal deformabilities ($\Lambda$) of compact stars \cite{Abbott2017,Abbott2020,Abbott2020ApJL}.
This data has the potential to revolutionize our understanding of the properties of matter under extreme conditions characterized by high density and pressure, where conventional laboratory experiments are not feasible.
Indeed, gravitational wave observations provide valuable constraints on the equation of state (EOS), which describes the relationship between pressure, density, and the structure of matter inside neutron stars. 
Analysis of GW170817 revealed a tight constraint on the tidal deformability parameter \cite{Abbott2018}, which disfavors very stiff EOSs \cite{IBombaci2021}. This feature assists in narrowing down the range of possible EOSs, offering significant constraints for theoretical models aiming to describe the interiors of neutron stars \cite{ITews2020}.
One of the challenges in investigating the internal structure of compact stars is determining the presence or absence of the quark matter phase. Theoretical Researchers such as Terazawa, Witten, and Bodmer have proposed the existence of quark stars based on the extreme conditions within compact stars \cite{Terazawa1989,Witten1984,ABodmer}. However, obtaining direct observational evidence for quark stars is challenging because of their similarities to neutron stars. Ref. \cite{Annala}, through independent model analysis, illustrates that when the conformal bound ($v^2/c^2\leq \frac{1}{3}$, where $v$ is the speed of sound and $c$ is the light speed) is not significantly violated, massive neutron stars are anticipated to contain substantial quark-matter cores. As a result, theoretically, compact stars are commonly classified into three categories:
i) Neutron stars, which consist entirely of hadronic matter \cite{Wiringa, Stone2007, Pearson, Negreiros,BordbarNS1,BordbarNS2,BordbarNS3,BordbarNS4},
ii) Hybrid stars, characterized by a quark core surrounded by hadronic layers \cite{Blaschke2001, Burgio2003, Alford2005, Pal2023, Rather2023, Li2023,BordbarNSQ1,BordbarNSQ2}, and iii)
strange quark stars (SQSs), comprised entirely of strange quark matter (SQM) \cite{Michel1988,Drago2001,Kurkela2010,Wang2019,Deb2021, Sedaghat2021,JSedaghat,BordbarSQS1}.

In this paper, our aim is to investigate the structural properties of SQSs using a Quantum Chromodynamics (QCD) perturbative model along with a density-dependent effective bag constant to account for non-perturbative contributions. To elaborate, our study shows that by modifying the perturbative EOS of SQM with a density-dependent effective bag constant, SQSs can exist with masses exceeding twice that of the Sun (such as PSR J0952-0607 \cite{J0952}, PSR J2215+5135 \cite{Linares2018}, PSR J0740+6620 \cite
{Cromartie2019}, and the secondary mass of GW190814 \cite{Abbott2020,ZMiao2021}), while satisfying gravitational wave constraints like tidal deformability. Additionally, this approach addresses thermodynamic considerations, including the stability condition of quark matter and  speed of sound  behavior. 
To describe SQM, the QCD perturbative model has advantages over other approaches such as the NJL model \cite{Mishustin, Hanauske, Buballa2005, Chu2016, ChengMingLi2020, sedaghat} and bag models \cite{Hua Li2010, MassIV, Deb2018, Podder2024} for the following reasons. 
1- Perturbation theory allows for systematic approximation by expanding physical quantities using a small parameter, typically a coupling constant. 2- The perturbation theory of QCD aligns with the Lagrangian of the standard model, making it well-suited for exploring the asymptotic behavior of physical quantities, such as the speed of sound. 3- The model relies on a renormalizable theory, crucial for guaranteeing finite and meaningful physical predictions. 4- The behavior of both the coupling constant and the strange quark mass is dependent on energy. This aspect is essential for accurately describing the EOS of SQM. However, at the energy scale present in compact stars, perturbative QCD  may not fully explain their structure. To address this, we introduce an effective bag parameter to incorporate non-perturbative contributions into the thermodynamic potential. Initially, we consider this non-perturbative component as a constant parameter. Subsequently, we define it as a  density-dependent function to investigate the structural properties of SQS. The structure of this paper is as follows: In the next section, we examine the running coupling and running strange mass using the latest Particle Data Group dataset. We then introduce a thermodynamic potential that includes both perturbative and non-perturbative components to calculate the EOS of SQM. Next, we treat the non-perturbative component as a constant parameter and determine its permissible values based on the stability condition of SQM. Following this, we obtain the EOSs of SQM and investigate its thermodynamic properties. We then calculate the structural properties of SQS to assess whether the results meet observational constraints. Subsequently, we repeat these steps by defining a density-dependent function for the non-perturbative component of the EOS and investigate its impact on the structural properties of SQS, considering observational constraints from different pulsars and the binaries of GW170817 and GW190814. We conclude the paper with some final remarks.\\

\section{Running coupling and running strange mass based on the latest Particle Data Group dataset} \label{mass and coupling}
\begin{figure}[h!]
	\centering
	\par
	\includegraphics[width=7cm]{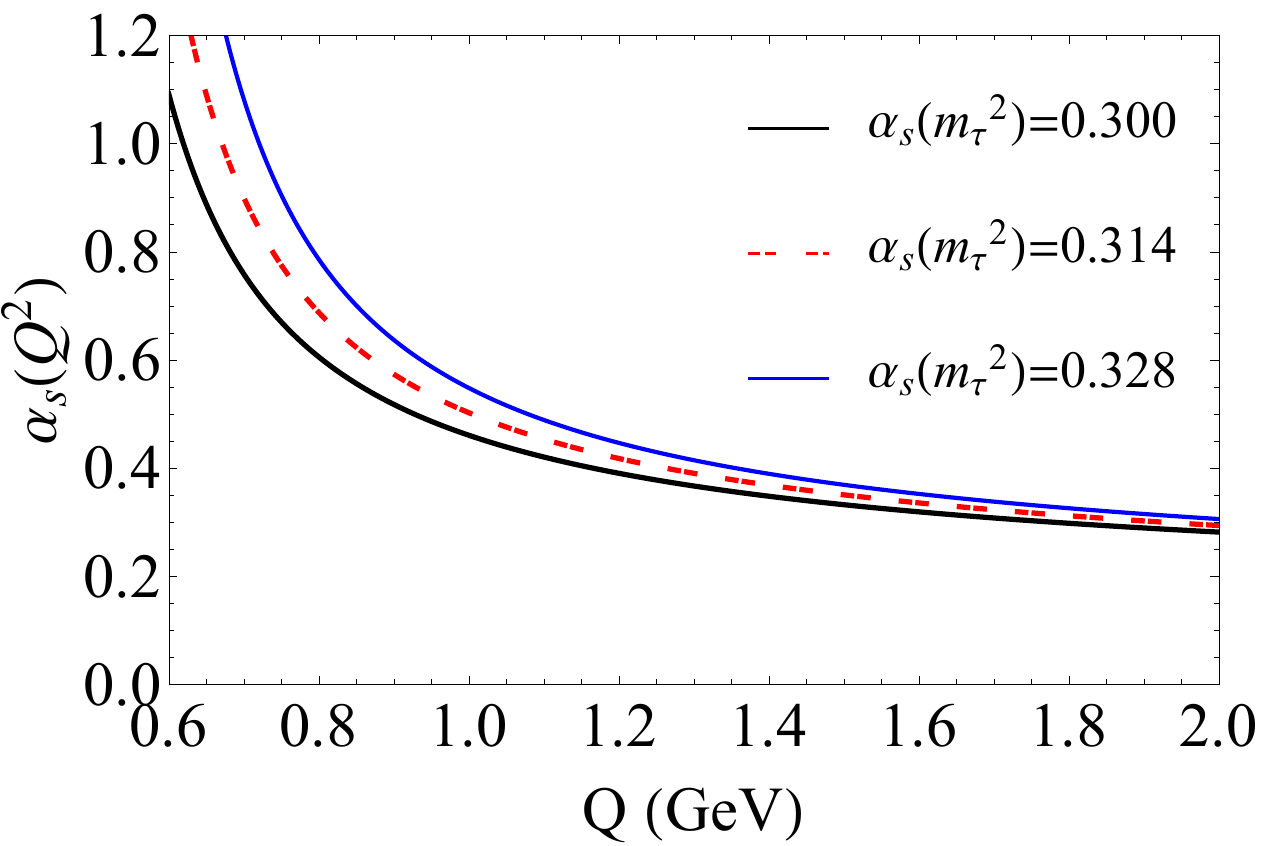}
	\caption{QCD running coupling constant versus energy.}
	\label{couplingdiagram}
\end{figure}

\begin{figure}[h!]
	\centering
	\par
	\includegraphics[width=7cm]{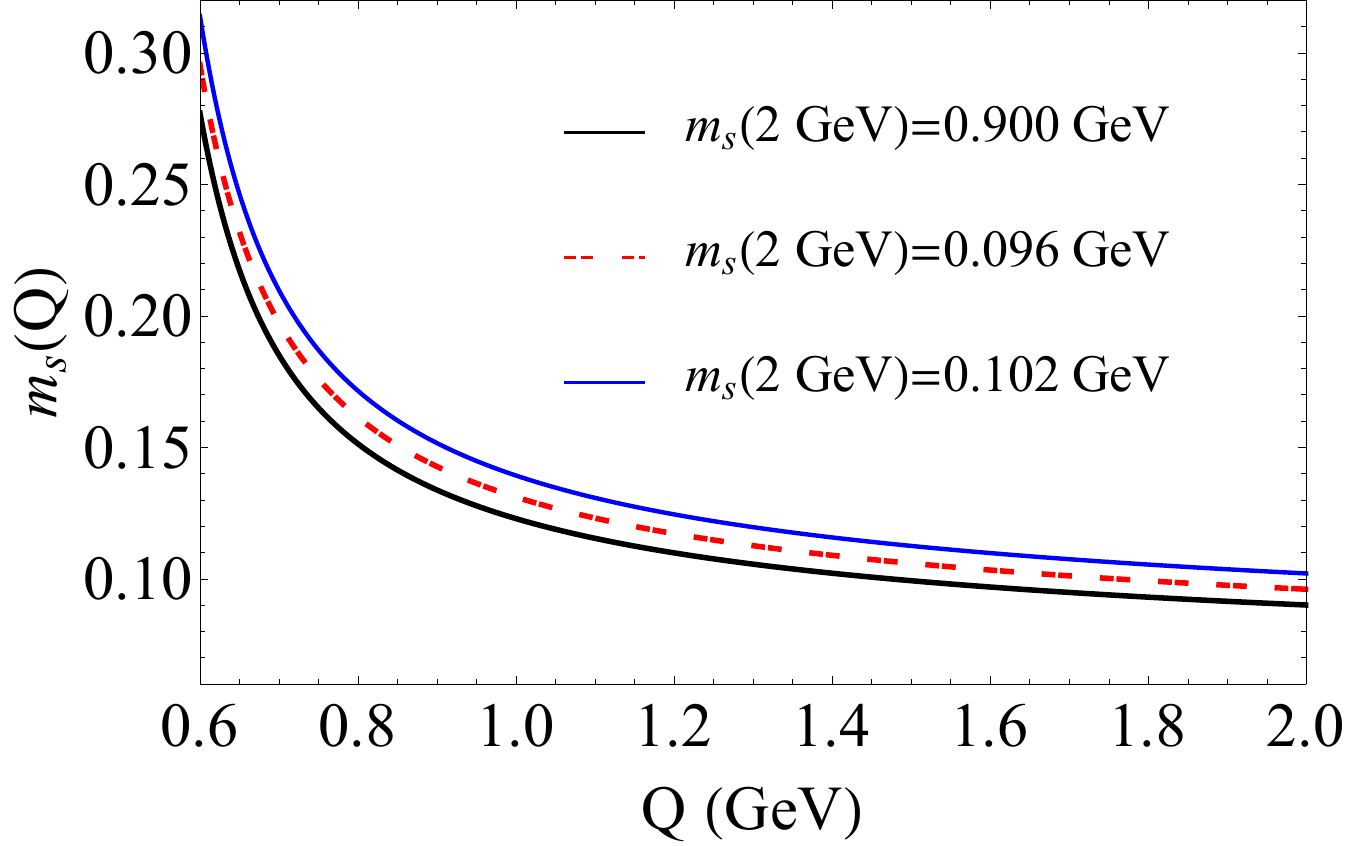}
	\caption{Running mass of the strange quark versus energy for different values of $m_s(2GeV)$.}
	\label{running mass}
\end{figure}

As commonly understood in perturbation theory, physical quantities are expanded in terms of the coupling constant. Therefore, our initial investigation focuses on how the QCD coupling constant,$\alpha_s$, evolves with energy.
The following relation describes the variation of $\alpha_s$ with respect to the renormalization scale, represented by $Q$ \cite{Vermaseren,Fraga2006}.  
\begin{equation} \label{QCD coupling}
	\alpha_s (Q) =\frac{4 \pi  \left(1-\frac{2\beta_1 \log (L)}{{\beta_0}^2 L}\right)}{\beta_0L},
\end{equation}
where $\beta_0= 11 - 2\frac{N_f}{3}$, $\beta_1=51 - 19\frac{N_f}{3}$,  $L=2\log(\frac{Q}{\Lambda_{\overline{MS}}})$, and $N_f$ is the number of flavors, which we set to $3$. $\Lambda_{\overline{MS}}$ represents the renormalization point in the minimal subtraction scheme, determined from the Particle Data Group 2023 dataset \cite{Workman} by ensuring that $\alpha_s(m_{\tau})=0.314^{+0.014}_{-0.014}$. $m_{\tau}$ is the mass of the tau particle, which is set to $1776.86 MeV$ \cite{Workman}. Fig. \ref{couplingdiagram} represents the behavior of $\alpha_s$ versus $Q$ for different values of $\alpha_s({m_{\tau}}^2)$. 
As depicted in Fig. \ref{couplingdiagram}, the uncertainty range of $\alpha_s$ diminishes as energy rises. Subsequently, in our calculations presented below, we adopt $\alpha_s(m_{\tau})$ as $0.314$.
This study concentrates on SQM, which includes up, down, and strange quark flavors at zero temperature and finite chemical potential. In this context, the masses of the up and down quarks are deemed negligible compared to that of the strange quark. At leading order, the energy dependence of the mass of the strange quark, denoted as $m_s(Q)$, can be described as follows \cite{Vermaseren,Fraga2006}.
\begin{equation}
	m_{s}(Q)=m_{s}(2GeV)\left[ \dfrac{\alpha _{s}(Q)}{\alpha _{s}(2GeV)}\right]
	^{\dfrac{\gamma _{0}}{\beta _{0}}},  \label{9}
\end{equation}%
where $\gamma _{0}$ is the anomalous dimension, equals $3\dfrac{N_{c}^{2}-1}{2N_{c}}$ in which $N_c=3$ is the number of colors. According to the latest Particle Data Group dataset, the mass of the strange quark at $2{GeV}$, denoted as $m_{s}(2{GeV})$, is $93.4^{+8.6}_{-3.4}{MeV}$. We consider the central value of this interval, $m_s(2{GeV}) = 96 {MeV}$, in our calculations.
The behavior of $m_s(Q)$ versus energy for different values of $m_s(2GeV)$ is shown in Fig. \ref{running mass}. Here, as illustrated in Fig. \ref{couplingdiagram}, it is evident that the uncertainty range decreases as energy increases. 
In the following section, we will calculate the thermodynamic potential of SQM at the leading order to establish the thermodynamic properties of the system.
\section{Thermodynamic potential}\label{e-p}
By using the thermodynamics potential of  the system, $\Omega $,  we can derive numerous properties of SQM, such as energy density, quark number density, pressure, speed of sound, adiabatic index, and additional characteristics. Because the interactions between quarks in QED are minimal compared to the QCD interaction, QCD emerges as the predominant force within the system.  We divide $\Omega$ into non-interacting and perturbative segments: i) the non-interacting part comprises  non-interacting quarks and electrons; ii) the perturbative segment involves QCD interactions among the quarks at the leading order. The perturbative component was previously derived from a two-loop Feynman diagram in Refs. \cite{Fraga2006} and \cite{Kurkela2010}. Here is the expression for the thermodynamic potential of a system comprising quark flavors with mass $m_f$ and chemical potential $\mu_f \ (f=u,d,s)$, along with electrons possessing a chemical potential $\mu_e$, up to the leading order \cite{Kurkela2010}. 
\begin{equation}
	-\frac{\Omega}{V}=\sum_{{N_f=1}}^{3}\left({\mathcal M}_1+\frac{{\mathcal M}_2\alpha_s(Q)}{4\pi}\right),
\end{equation}
where ${\mathcal M}_1$ and $\frac{{\mathcal M}_2\alpha_s(Q)}{4\pi}$ are  non-interacting and perturbative parts, respectively. ${\mathcal M}_1$ and ${\mathcal M}_2$ are presented as follows
\begin{equation}
	{\mathcal M}_1=
	\frac{N_c {\mu_f}^4}{24\pi^2}\bigg\{2\hat{u_f}^3-3z_f \hat{m_f}^2\bigg\} +\frac{{\mu_e}^4}{12\pi^2},\label{m1}
\end{equation}
\begin{equation}
	{\mathcal M}_2= \frac{d_A {\mu_f}^4}{4\pi^2}\Bigg\{-6z_f \hat{m_f}^2 \ln\frac{Q}{m} +2\hat{u_f}^4 - 4z_f \hat{m_f}^2 -3{z_f}^2 \Bigg\}, \label{m2}
\end{equation}
where $\hat{u_f}\equiv(\sqrt{{\mu_f}^2-m_f^2})/\mu_f$, $\hat{m_f}\equiv m_f/\mu_f$, $z_f \equiv \hat{u_f}-\hat{m_f}^2\,\ln\bigg[\frac{1+\hat{u_f}}{\hat{m_f}}\bigg]$ and $d_A\equiv {N_c}^2-1$.  Due to beta equilibrium, the following relations are established between $\mu_u$, $\mu_d$, $\mu_s$, and $\mu_e$. 
\begin{eqnarray}
	\mu _{s}&=&\mu _{d}\equiv \mu ,\nonumber\\\mu _{u}&=&\mu -\mu _{e}.  \label{24}
\end{eqnarray}
It is worth noting that, based on the results obtained for a stable SQM from \cite{Kurkela2010}, we set $Q=\frac{4}{3}(\mu_u+\mu_d+\mu_s)$. Now, we can derive the pressure using the relation $P=-B-\frac{\Omega }{V}$, where the parameter $B$ accounts for non-perturbative effects not encompassed by the perturbative expansion \cite{Kurkela2010,Sedaghat2021,JSedaghat}. The values of $B$ are determined to ensure that the total pressure at the star's surface equals zero  \cite{Kurkela2010,Sedaghat2021,JSedaghat,sedaghat}. If we exclude the perturbative part, the parameter $B$ is simply the bag constant. However, when we include the perturbative part, its value differs from those in the MIT bag model. Therefore, we define it as an effective bag constant. The energy density, $\epsilon$, is given as
\begin{equation}\label{EOS}
	\epsilon=\sum_{{N_f}}\mu_fn_f+\mu_en_e-P,
\end{equation}
where $n_f$ represents the quark number density with flavor $f$, defined as $\frac{\partial P}{\partial \mu_f}$, while $n_e$ denotes the electron number density, derived from the relation $n_{e}=\frac{\mu_{e}^{3}}{(3\pi^2 )}$. Relation (\ref{EOS}) states the correlation between pressure and energy density, commonly known as the EOS. Understanding the EOS allows us to explore the behaviors of various thermodynamic quantities, such as the speed of sound or the adiabatic index, and to obtain the structural characteristics of the SQS. However, before proceeding with such investigations, it is crucial to verify the stability condition of SQM. In the following section, we will elucidate this condition and determine the range of values for parameter $B$ that satisfy it. 
\section{Stability of SQM}\label{sc}
\begin{figure}[h!]
	\centering
	\includegraphics[width=7cm]{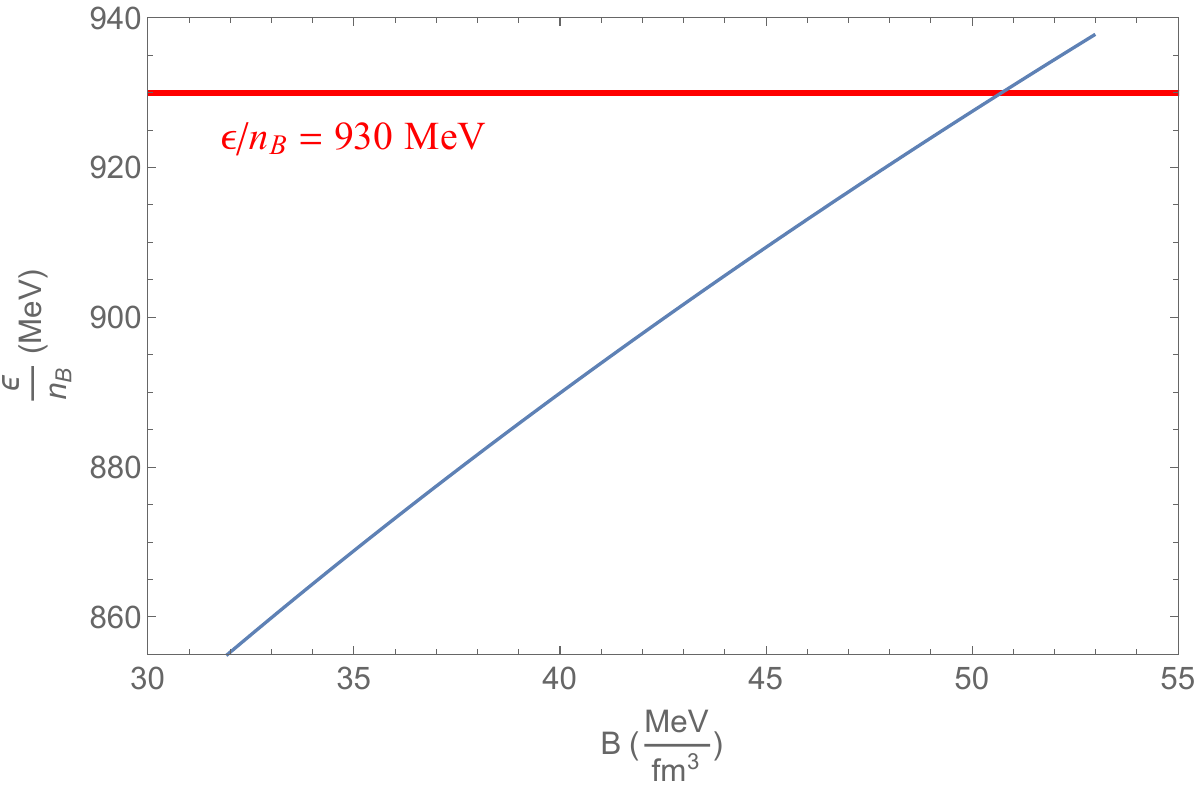}
	\caption{Energy per baryon for different values of $B$.}	\label{stabilitycondition}
\end{figure}

As previously mentioned, the parameter $B$ is included in the perturbative EOS to account for non-perturbative effects. Initially, we assume that $B$ is a constant parameter and determine the maximum $M_{TOV}$ of the SQS by considering the conditions of thermodynamic stability and observational constraints. Next, we introduce a density-dependent function for $B$ and repeat the process to evaluate its impact on the results. We will show that with a constant $B$, the EOSs cannot describe compact objects with $M_{TOV} > 2.03M_{\odot}$, while considering $B$ as a  density-dependent function significantly changes the results.
It is worth noting that the parameter $B$ also appears in other models, such as the NJL model and the MIT bag model. The range of $B$ varies between models, depending not only on the specific model used but also on factors such as the number of quark flavors, the coupling constant, the mass of the quarks, and other variables. For example, in Ref. \cite{Stergioulas}, the authors use the MIT bag model and consider $m_s = 0$ to constrain $B$ to the range $58.9 \, \frac{\text{MeV}}{\text{fm}^3} < B < 91.5 \, \frac{\text{MeV}}{\text{fm}^3}$. In Ref. \cite{Farhi1984}, the authors assume $m_s = 150 \, \text{MeV}$ and find that $B$ lies in the range $56-78 \, \frac{\text{MeV}}{\text{fm}^3}$. In Ref. \cite{Aziz2019}, the researchers analyze 20 compact stars and determine that $B$ ranges from $41.58$ to $319.31 \, \frac{\text{MeV}}{\text{fm}^3}$. In Ref. \cite{Zhou2018}, the authors use the constraint $\Lambda_{1.4\textup{M}_\odot}<800$ to narrow down $B$ to the range $42 \, \frac{\text{MeV}}{\text{fm}^3} < B < 52 \, \frac{\text{MeV}}{\text{fm}^3}$. In this section, we consider a constant $B$ and investigate the stability condition of SQM. We aim to determine the range of $B$ values that meet this stability criterion. If SQM exists, it represents the true ground state of QCD, indicating that the minimum energy per baryon in SQM at zero pressure is lower than that of the most stable nuclear matter. Consequently, if $n_B=\frac{n_u+n_d+n_s}{3}$ represents the baryon number density, we must impose the condition  $\frac{\epsilon}{n_B}<930MeV$ \cite{Weber2005}.
At baryon densities below $1.1n_{sat}$, matter resides in the hadronic phase, where $n_{sat}$ denotes the nuclear saturation density, set at 0.16. Assuming $n_B > 1.1n_{sat}$ and enforcing charge neutrality through the equation $\frac{2}{3}n_u-\frac{1}{3}n_d-\frac{1}{3}n_s-n_e=0$, we determine the range of the effective bag constant, $B$, in which the stability condition is fulfilled. Fig. \ref{stabilitycondition} shows the energy per baryon in terms of $B$. The red line corresponds to the value $\frac{\epsilon}{n_B}=930MeV$. We observe that for values of $B$ greater than $51\frac{MeV}{fm^3}$, the stability condition is not met. Now, we have reached a point where we can determine the EOS of SQM and investigate the behavior of the speed of sound and the adiabatic index. In the next section, we will compute the EOS for various permissible values of $B$. 

\section{Thermodynamic properties of SQM at various constant $B$ values}
\begin{figure}[h!]
	\centering
	\par
	\includegraphics[width=7cm]{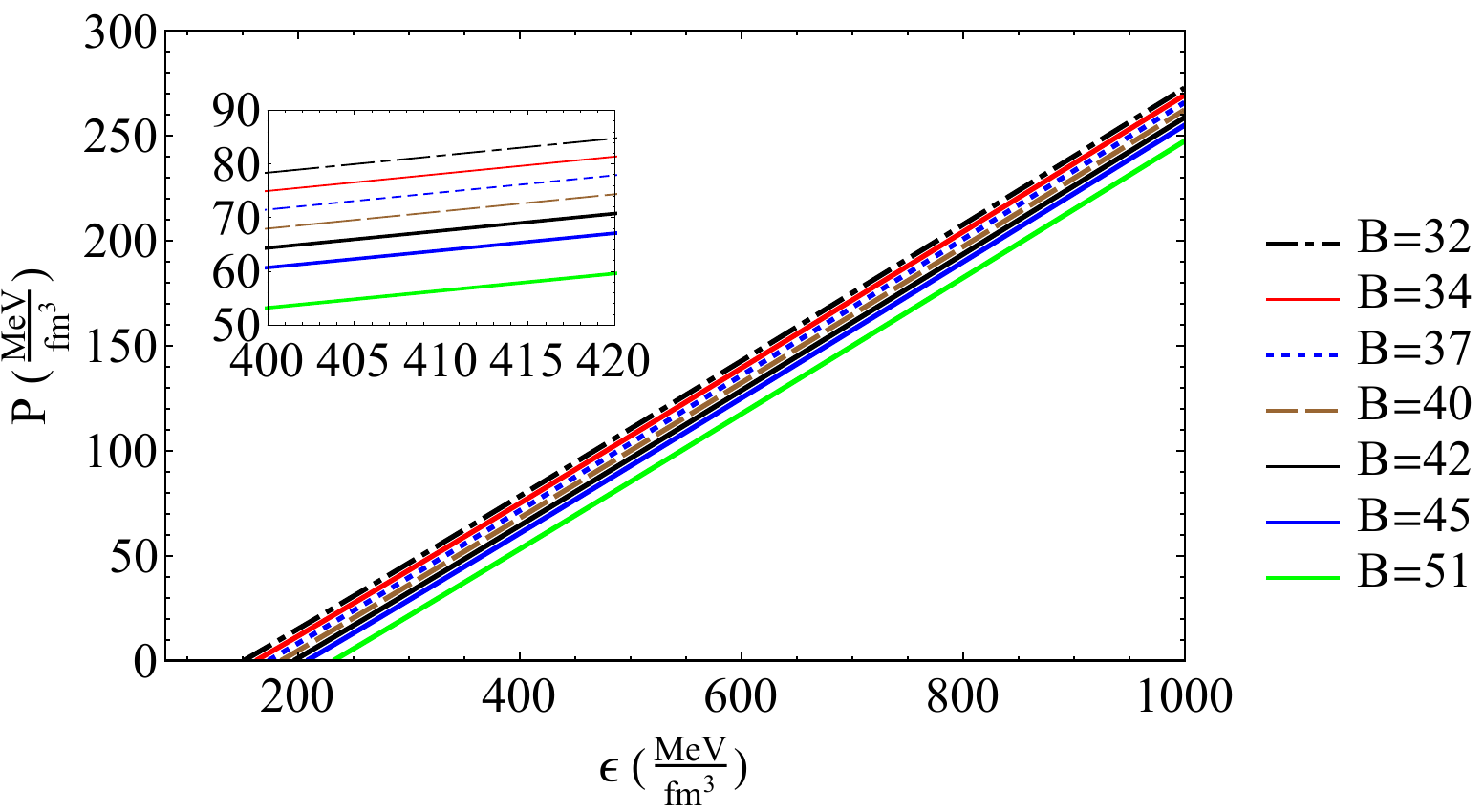}
	\caption{EOS of SQM for different values of $B$.}\label{EOSs}
\end{figure}

\begin{figure}[h!]
	\centering
	\par
	\includegraphics[width=7cm]{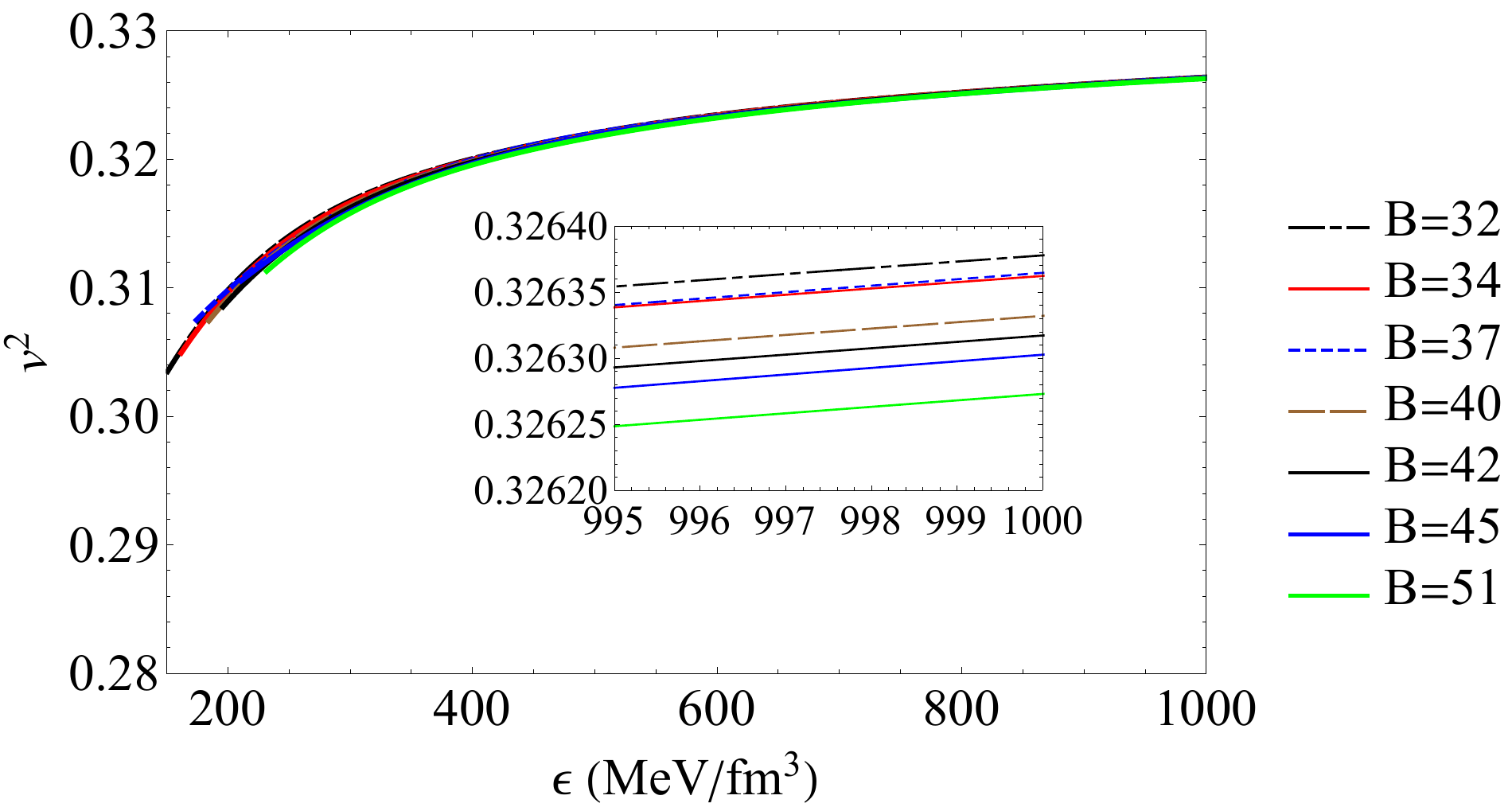}
	\caption{speed of sound (in the unit of light speed $c$) for different values of $B$.}\label{sound1}
\end{figure}

\begin{figure}[h!]
	\centering
	\par
	\includegraphics[width=7cm]{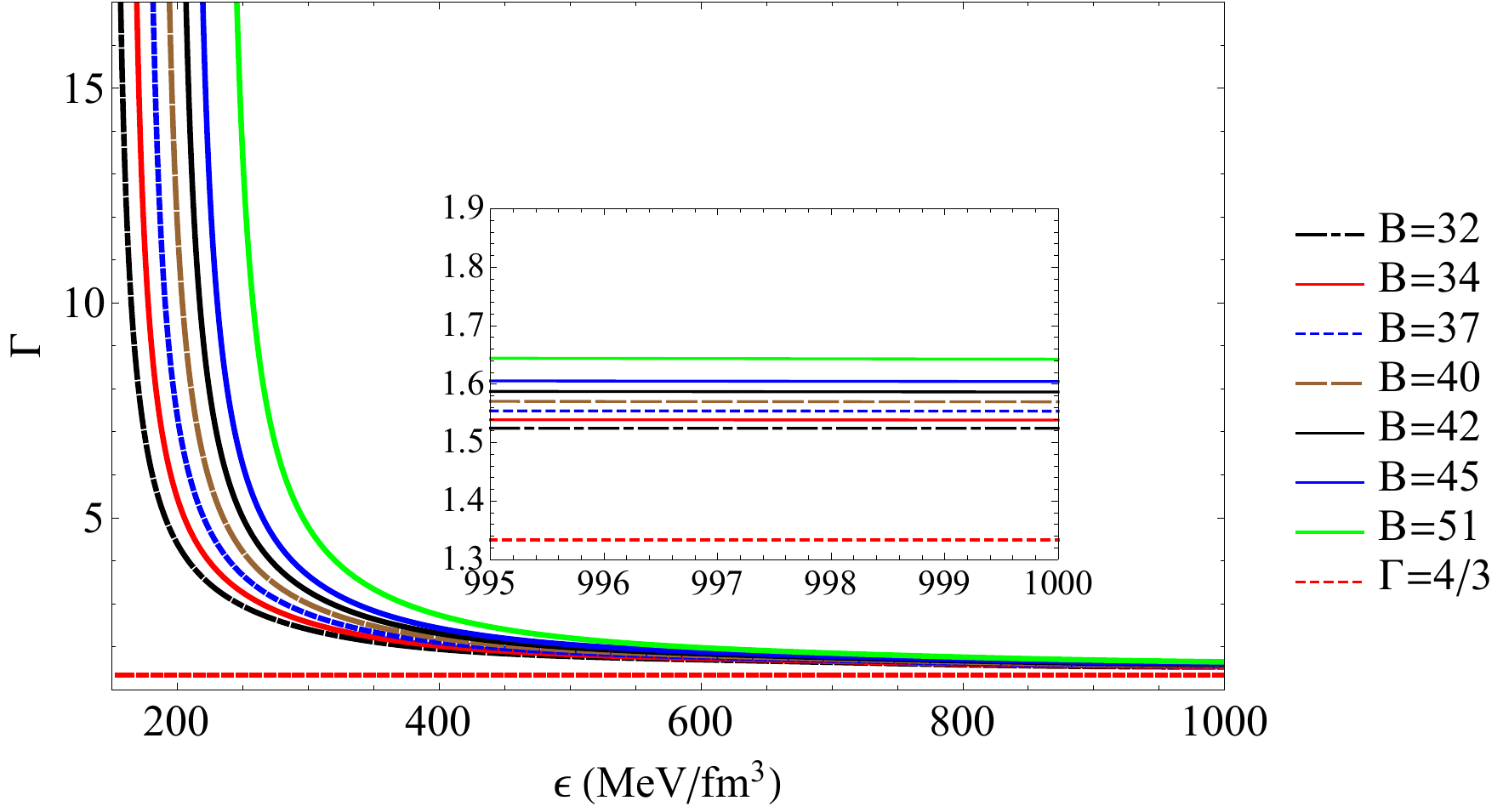}
	\caption{Adiabatic index for SQM at different values of $B$.}\label{adia1}
\end{figure}

By calculating pressure and energy density (as discussed in section \ref{e-p}) while considering beta equilibrium and charge neutrality conditions, we derive the EOSs for various permissible values of $B$ (as discussed in section \ref{sc}). By knowing the EOS, we can determine other thermodynamic properties, such as the adiabatic index and the speed of sound. Fig. \ref{EOSs} illustrates that increasing $B$ softens the EOS. This characteristic influences the structural properties of SQS. In the next section, we will observe that increasing $B$ results in a decrease in both the maximum gravitational mass and tidal deformability. 
To ensure causality, the speed of sound must be less than the speed of light in a vacuum ($c$). Fig. \ref{sound1} illustrates that this requirement is met satisfactorily for all diagrams corresponding to each value of $B$. Additionally, Fig. \ref{sound1} demonstrates that the results converge to the value of $v^2/c^2=\frac{1}{3}$ at high energy densities, corresponding to a non-interacting relativistic gas.
It is recognized that the adiabatic index, $\Gamma =\frac{dP}{d\epsilon} \dfrac{(P+\epsilon )}{P}$, must exceed $4/3$ to meet dynamical stability \cite%
{Chandrasekhar1964,Bardeen1966,Kuntsem1988,Mak2013}. The adiabatic index is depicted as a function of energy density in Fig. \ref{adia1}. As evident from this figure, the condition $\Gamma > 4/3$ is established for all diagrams.
In the next section, we will proceed with the calculation of the structural features of SQS.  Following the observational constraints, we will determine the maximum allowed mass of SQS. 
\section{Structural properties of SQS at various constant $B$ values}
In this section, we explore the structural properties of SQS using the EOSs obtained in the previous section, with a focus on gravitational mass, corresponding radius, and tidal deformability. By considering the constraint on $\Lambda$ from GW170817, we specifically investigate the maximum $M_{TOV}$ provided by the EOSs with constant values of $B$. In the following, we first derive the mass as a function of central energy density and the radius of the star, and then we explain its dependence on $\Lambda$.

\subsection{Mass and radius} 
\begin{figure}[h!]
	\centering
	\par
	\includegraphics[width=7cm]{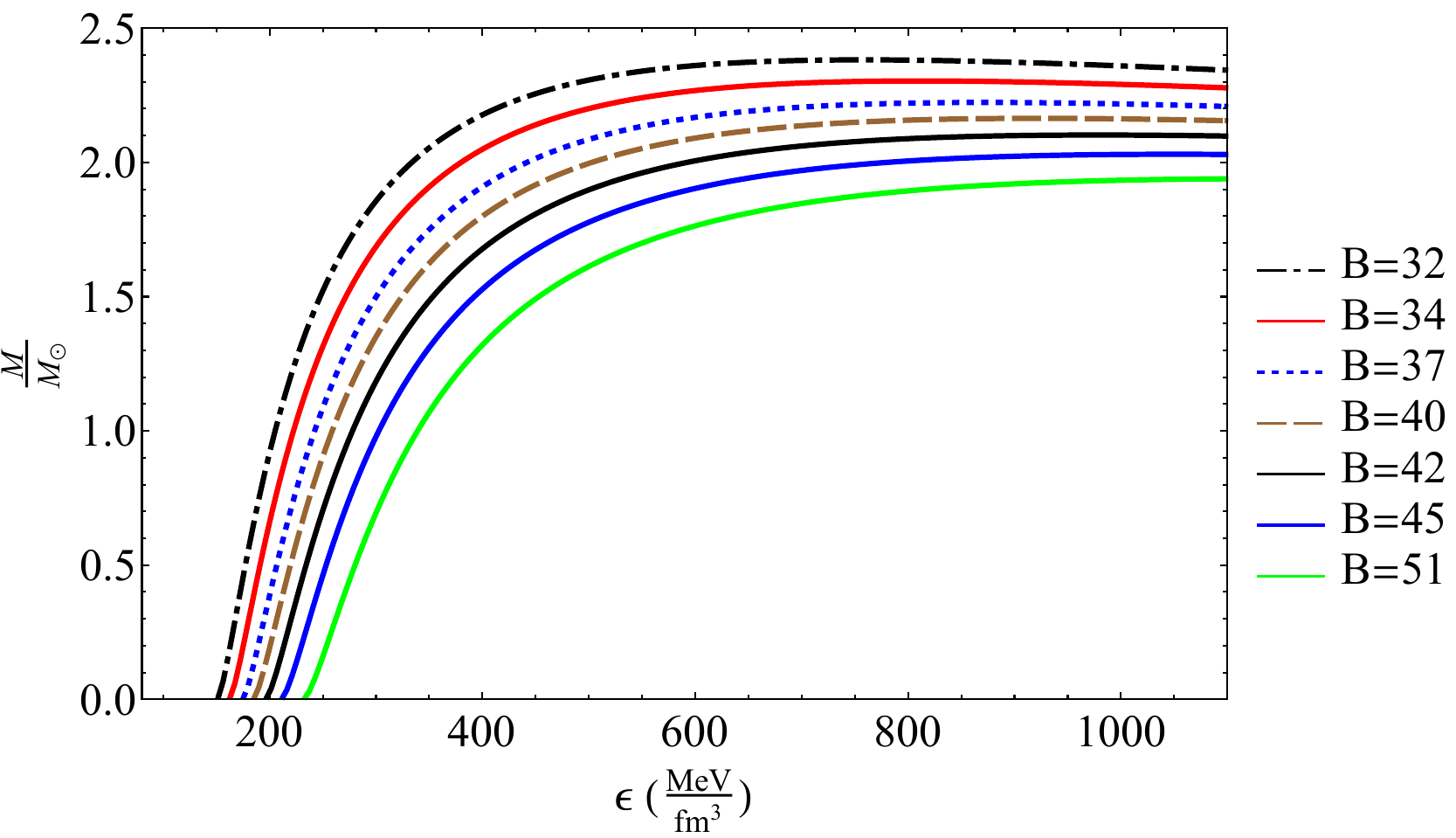}
	\caption{Mass versus central energy density for different values of $B$.}
	\label{Mediagrams}
\end{figure}

\begin{figure}[h!]
	\centering
	\par
	\includegraphics[width=7cm]{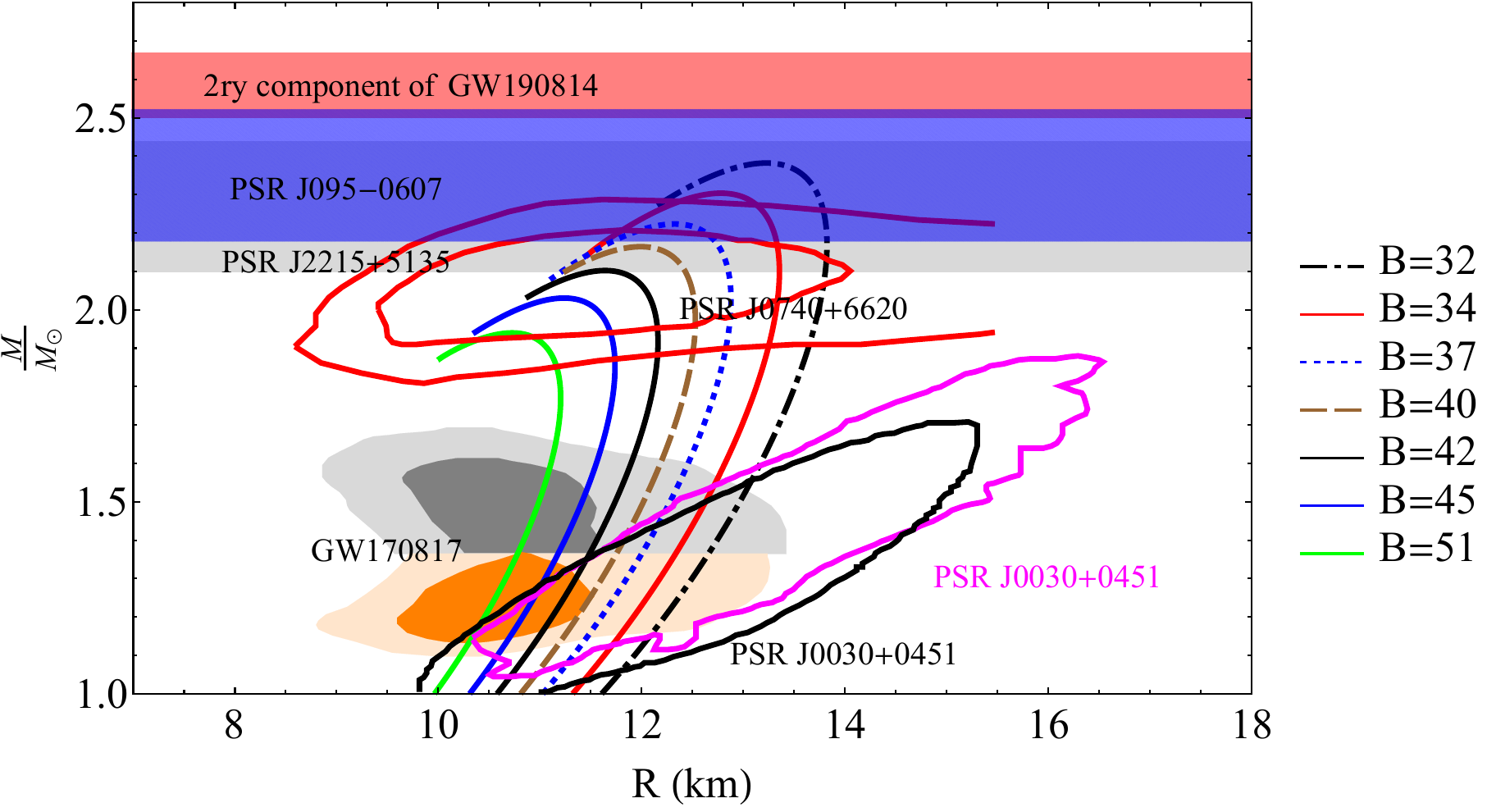}
	\caption{M-R diagram for different values of $B$.}
	\label{MRdiagrams}
\end{figure}

To obtain the mass in terms of energy density as well as the mass in terms of radius, we employ the TOV equation \cite{Tolman1939,Oppenheimer1939}. 
\begin{eqnarray}
	\frac{dP(r)}{dr} &=&\dfrac{\left[ P(r)+\epsilon (r)\right] \left[ M(r)+4\pi
		r^{3}P(r)\right] }{r\left( 2M(r)-r\right) },\label{TOV}
\end{eqnarray} 
where, $r$ denotes the radial coordinate of the star. This equation is solved concurrently with the equation $M(r)=\int_{0}^{R}4\pi r^{2}\epsilon(r)dr$.
To solve these equations, we must consider central pressures in line with the perturbative EOSs. It is worth noting that the mass of the star at its center is zero. Subsequently, we proceed to solve the equations until we attain the star's surface, where the pressure equals zero. The outcome yields both the mass and corresponding radius of the star, as well as the relationship between mass and central pressure or central energy density. This iterative process is carried out for various values of central pressures permitted by the EOSs. Our results for gravitational mass versus central energy density is shown in Fig. \ref{Mediagrams}. The trend observed in the mass depicted in Fig. \ref{Mediagrams} signifies the maximum gravitational mass of SQS, denoted as $M_{TOV}$. {In fact for stability of the star, the condition $\frac{\partial M}{\partial \epsilon_c}>0 $ should be imposed, where $\epsilon_c$ denotes the energy density at the center of the star. For more discussion see \cite{Tangphati2021a,Tangphati2021b}.} Analyzing this figure allows us to ascertain the range of central energy densities for SQS. 
The mass behavior in terms of radius ($M-R$ relation) is illustrated in Fig. \ref{MRdiagrams} for various EOSs. Fig. \ref{MRdiagrams} displays various colored regions. The gray and orange areas show the mass and radius range obtained from GW170817. The solid black and pink curves correspond to PSR J0030+0451 \cite{Miller2019,Riley2019}. The red curve represents PSR J0740+6620 \cite
{Cromartie2019}. The gray bar denotes the mass range of PSR J2215+5135 \cite{Linares2018}. The blue and red bars indicate the mass range of PSR J0952-0607 \cite{J0952} and the lower mass of the binary GW190814 \cite{Abbott2020,ZMiao2021}, respectively. As evident from the figure, the obtained results cover the mass and radius regions detected by the NICER and LIGO detectors very well. However, we are investigating whether these outcomes meet the constraints derived from LIGO for $\Lambda$.
\subsection{Tidal deformability}
\begin{figure}[h!]
	\centering
	\par
	\includegraphics[width=7cm]{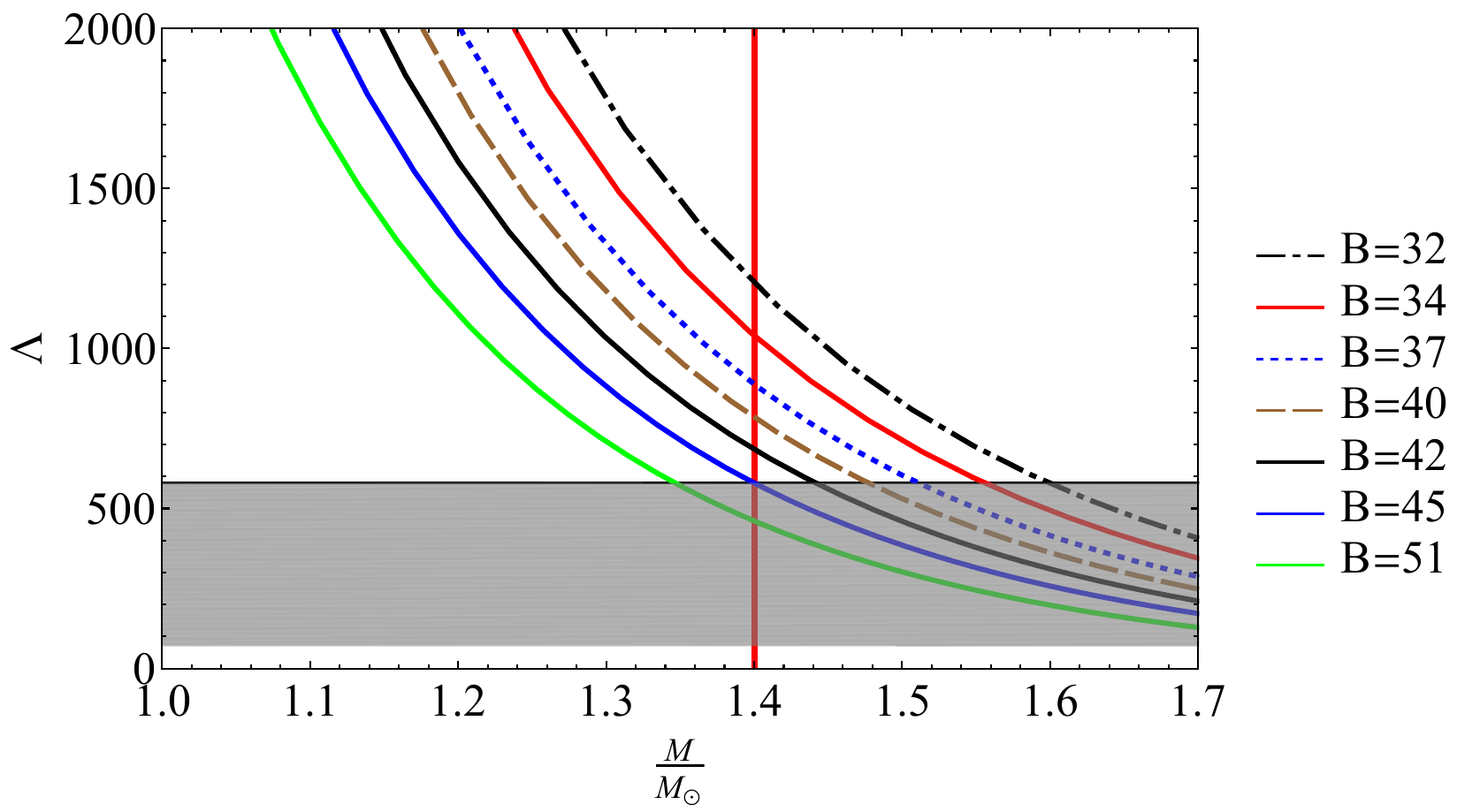}
	\caption{$\Lambda-M$ diagram for different values of $B$.}
	\label{tidal}
\end{figure}

To determine the tidal deformability, we must derive the metric function $H(r)$, which satisfies to the following equation \cite{Hinderer,ChengMingLi2020,sedaghat},
\begin{eqnarray}
		\frac{d\beta}{dr} &=& 2(1-2\frac{m_r}{r})^{-1}H\{-2\pi[5\epsilon+9p+f(\epsilon+p)]\nonumber\\
	&+&\frac{3}{r^2}+2(1-2\frac{m_r}{r})^{-1}(\frac{m_r}{r^2}+4\pi rp)^2\}\nonumber\\
	&+&\frac{2\beta}{r}(1-2\frac{m_r}{r})^{-1}\{\frac{m_r}{r}+2\pi r^2(\epsilon-p)-1\}\,\label{HbetaEq}
\end{eqnarray}
In this relation, $\beta = \frac{dH}{dr}$ and $f$ represents $\frac{d\epsilon}{dp}$.  The parameter $\Lambda $ is then calculated using the formula:
\begin{equation}
	\Lambda =\frac{2}{3}k_{2}R^{5},
\end{equation}
where $k_{2}$ is the dimensionless tidal Love number for $l=2$. The expression for $k_{2}$ is:
\begin{align}
	k_{2}& =\frac{16\sigma ^{5}}{5}(1-2\sigma )^{2}\left[ 1+\sigma (y-1)-\frac{y%
	}{2}\right]  \notag \\
	& \times \left\{ 12\sigma \left[ 1-\frac{y}{2}+\frac{\sigma (5y-8)}{2}\right]
	\right.  \notag \\
	& +4\sigma ^{3}\left[ 13-11y+\sigma \left( 3y-2\right) +2\sigma ^{2}\left(
	1+y\right) \right]  \notag \\
	& +\left. 6(1-2\sigma )^{2}\left[ 1+\sigma (y-1)-\frac{y}{2}\right] \ln
	\left( 1-2\sigma \right) \right\} ^{-1/2},  \label{tln}
\end{align}
where $\sigma$ is defined as $M/R$, and $y$ is given by $y = \frac{R \beta (R)}{H(R)} - \frac{4\pi R^{3} \epsilon_{0}}{M}$, in which $\epsilon_{0}$ denotes the energy density at the surface of the SQS \cite{ChengMingLi2020,sedaghat}.
By concurrently solving equations \ref{TOV} and \ref{HbetaEq}, along with $\frac{dM}{dr} = 4\pi r^{2} \epsilon$, we derive the relationship representing $\Lambda$ as a function of mass. Now, we examine our results based on the constraint on $\Lambda$ derived from the binary systems GW170817. 
Fig. \ref{tidal} represents $\Lambda-M$ diagram for different EOSs. As illustrated from the figure, the EOSs with $B<45\frac{MeV}{fm^3}$ fail to satisfy the constraint $\Lambda_{1.4M\odot}<580$. Table \ref{results1} displays our results regarding the structural properties of SQS for various permissible values of $B$. From the table, it is evident that the resulting EOSs cannot lead to SQSs with $M_{TOV}>2.03M_{\odot}$ while meeting the constraint $70<\Lambda_{1.4M\odot}<580$.  Up to this point, all results of the structural properties of SQS have relied on the assumption that the bag parameter remains constant across all densities of the SQS. However, for results of higher accuracy, it is recommended to incorporate a bag parameter that varies with density into the calculations. In the following section, we will examine how taking into account a density-dependent bag parameter affects both the thermodynamic properties of SQM and the structural features of SQS. 
\begin{table}[h!]
	\caption{{\protect\small {Structural properties of SQS for different values of $B$.}}}
	\label{results1}\centering
	\par
	\begin{adjustbox}{width=.42\textwidth}
		\begin{tabular}{|c|c|c|c|c|c|c|}
			\hline
			$B(MeV/fm^3)$ & $\Lambda_{1.4\textup{M}_\odot}$& $R(km)$ & $M_{TOV}(\textup{M}_\odot)$\\	\hline
			$32$ & 1199.73 & 12.93 & 2.37 \\ \hline
			$34$ &1038.66 & 12.63 & 2.30 \\ \hline
			$37$ & 877.24 & 12.41 & 2.22 \\ \hline
			$40$ & 784.90 & 11.95 & 2.16\\ \hline
			$42$ & 684.84& 11.56 & 2.10\\ \hline
			$45$ & 576.35& 11.19 & 2.03\\ \hline
			$51$ & 459.86 & 10.72 & 1.94 \\ \hline
		\end{tabular}
	\end{adjustbox}
\end{table}
\section{Density dependent bag constant} \label{DensityDBag}
\begin{figure}[h!]
	\centering
	\par
	\includegraphics[width=7cm]{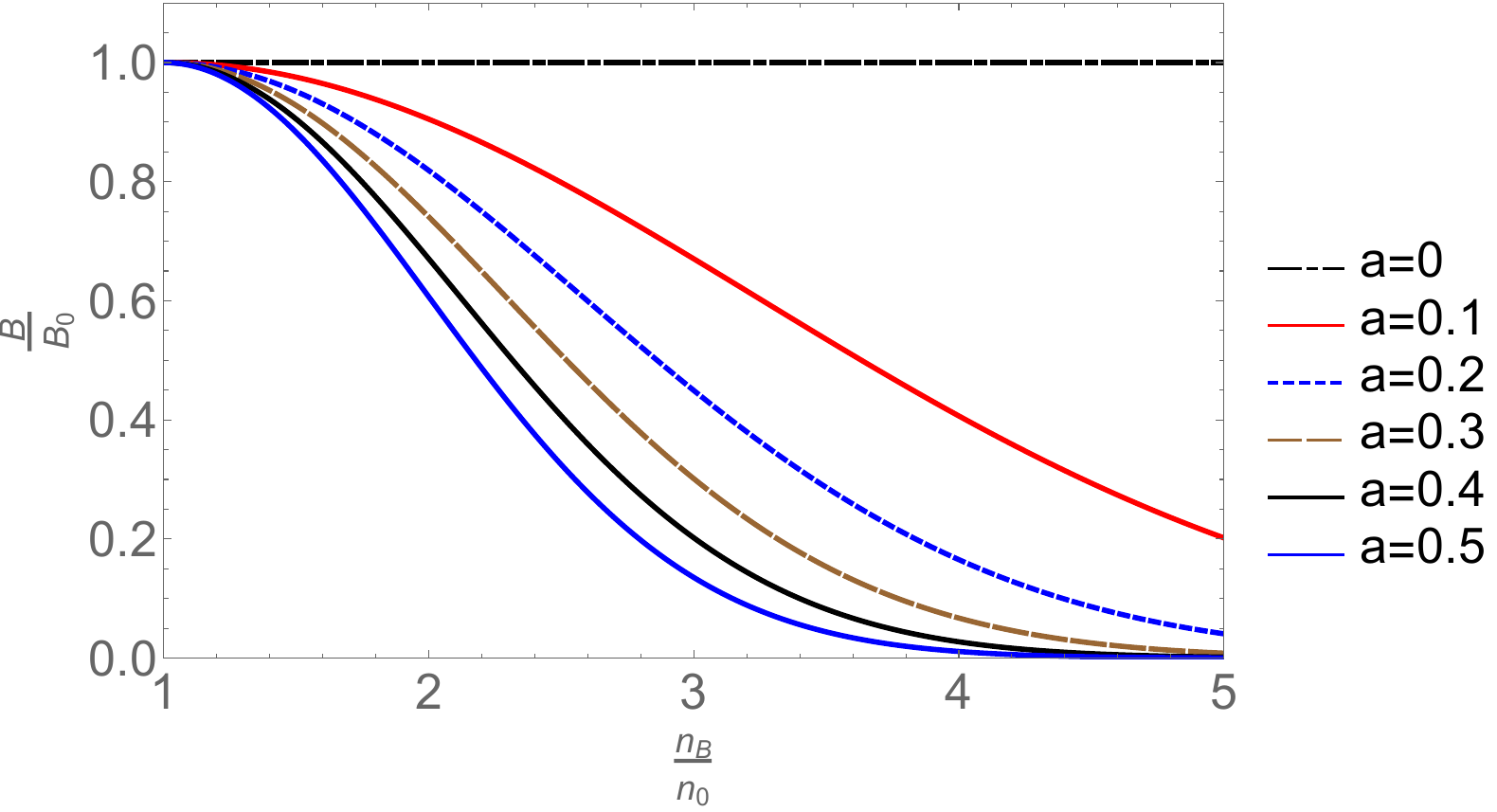}
	\caption{Density-dependent bag constant versus baryon number density for different values of $a$.}
	\label{bag2}
\end{figure}

So far, we have used a QCD perturbative model along with a constant $B$ to account for non-perturbative contributions in obtaining the EOS of SQM.
In the previous section, we demonstrated that the obtained EOSs cannot lead to a SQS with $M_{TOV} > 2.03M_\odot$ while also meeting the constraint $70 < \Lambda_{1.4M_\odot} < 580$. This limitation for $M_{TOV}$ might arise from the assumption of a constant $B$ in the EOS of SQM. Given that density increases from the surface to the center of the star, using a constant $B$ might not be an accurate approximation. As we know, with increasing density, the QCD coupling constant decreases, leading to a corresponding decrease in the bag constant. Therefore for more accurate description the parameter $B$ needs to be density-dependent \cite{G.F Burgio2002,Prasad2004,Bordbar2012,Pal2023,Podder2024}. Our model for density-dependent $B$ is represented as follows:
\begin{equation}
B=B_0e^{-a(\frac{n_B}{n_0}-1)^2}, \label{DDbag}
\end{equation}
where $B_0$ and $n_0$ are defined as the bag pressure and baryon number density at the surface of the star, respectively. The values of $B_0$ are determined by the stability condition of SQM at zero pressure, which corresponds to the pressure at the surface of the star. In a prior analysis detailed in section \ref{sc}, we derived $B_0$ values that comply with this stability criterion. The parameter $a$ dictates the variation of $B$ relative to density. A higher $a$ results in a quicker decrease of the bag constant, while a smaller $a$ leads to slower changes in $B$. Consequently, when $a$ is set to zero, we return to the constant $B$ approximation, which we previously investigated. For a more clear depiction, Fig. \ref{bag2} exhibits the ratio $\frac{B}{B_0}$ across different values of $a$. The figure demonstrates that this ratio decreases as $a$ increases. {It should be noted that the behavior of $B$ inside the star is uncertain. However, the proposed Gaussian form appears to align well with the asymptotic properties of QCD. In the following we will see that by selecting this type of function for $B$, we derive EOSs that not only meet the criteria for dynamic stability but also ensure appropriate sound speed behavior in ultra relativistic limit. Furthermore, the EOSs lead to SQSs with masses reaching up to $2.6 M_\odot$ while simultaneously satisfy observational constraints on mass, radius, and tidal deformability.} In the following section we choose different values of $a$ and investigate its effect on the behavior of the speed of sound and consequently structural properties of SQS.
\section{Thermodynamic properties of SQM for  density-dependent $B$}
\begin{figure}[h!]
	\centering
	\begin{subfigure}{0.45\textwidth}
		\centering
		\includegraphics[width=\textwidth]{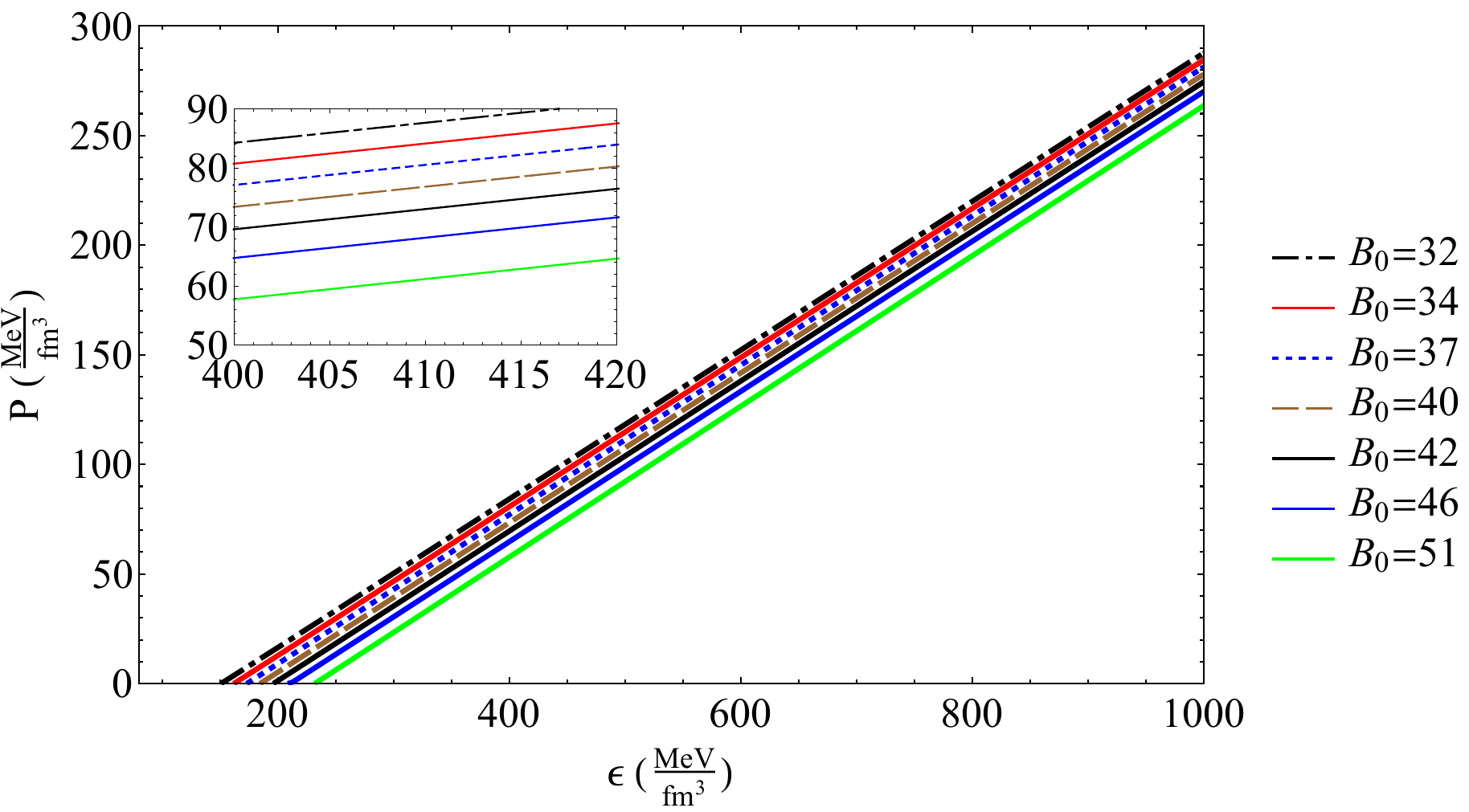}
		{a=0.1}
	\end{subfigure}
	\hfill
	\begin{subfigure}{0.45\textwidth}
		\centering
		\includegraphics[width=\textwidth]{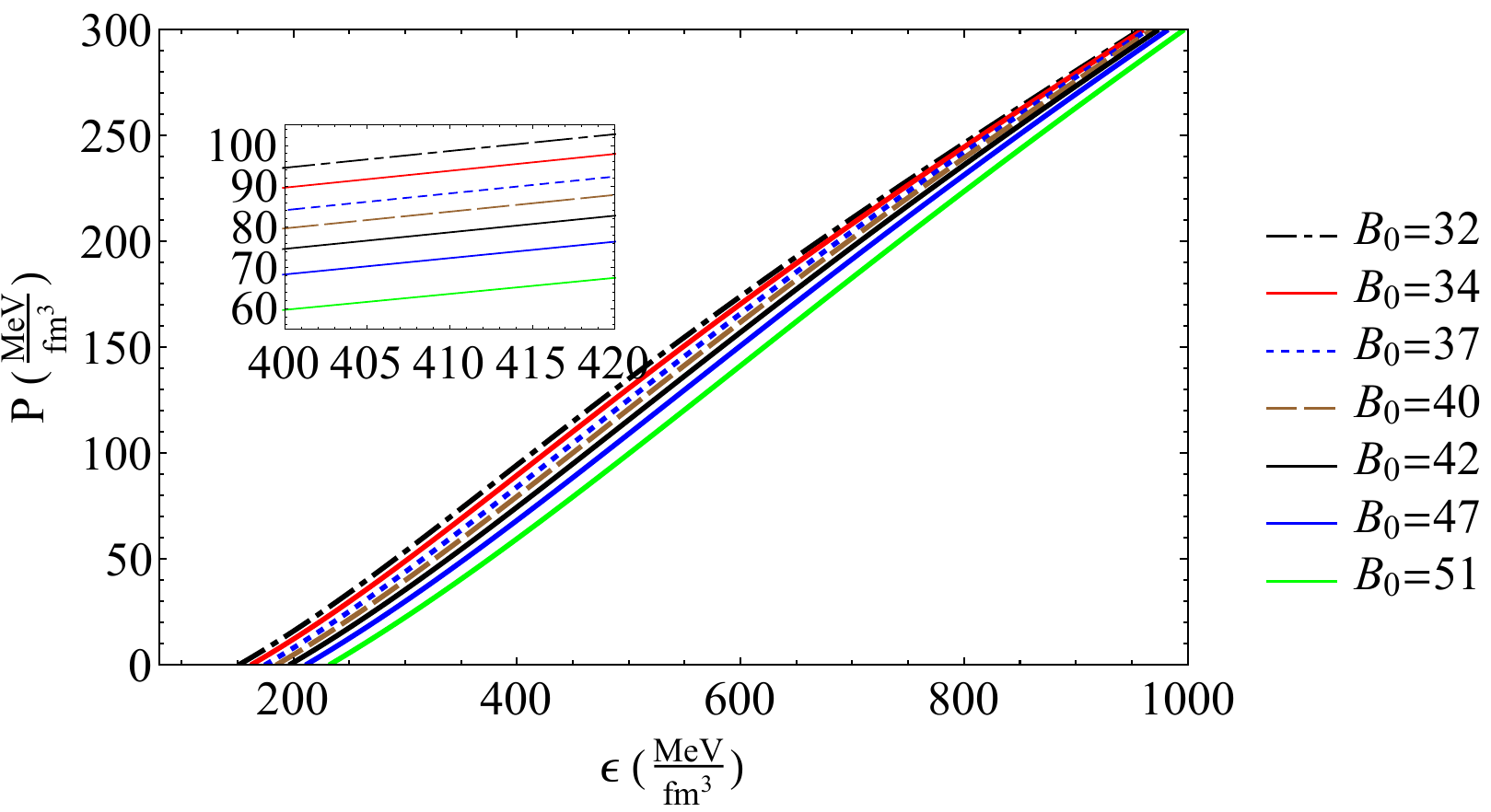}
		{a=0.2}
	\end{subfigure}
	\hfill
	\begin{subfigure}{0.45\textwidth}
		\centering
		\includegraphics[width=\textwidth]{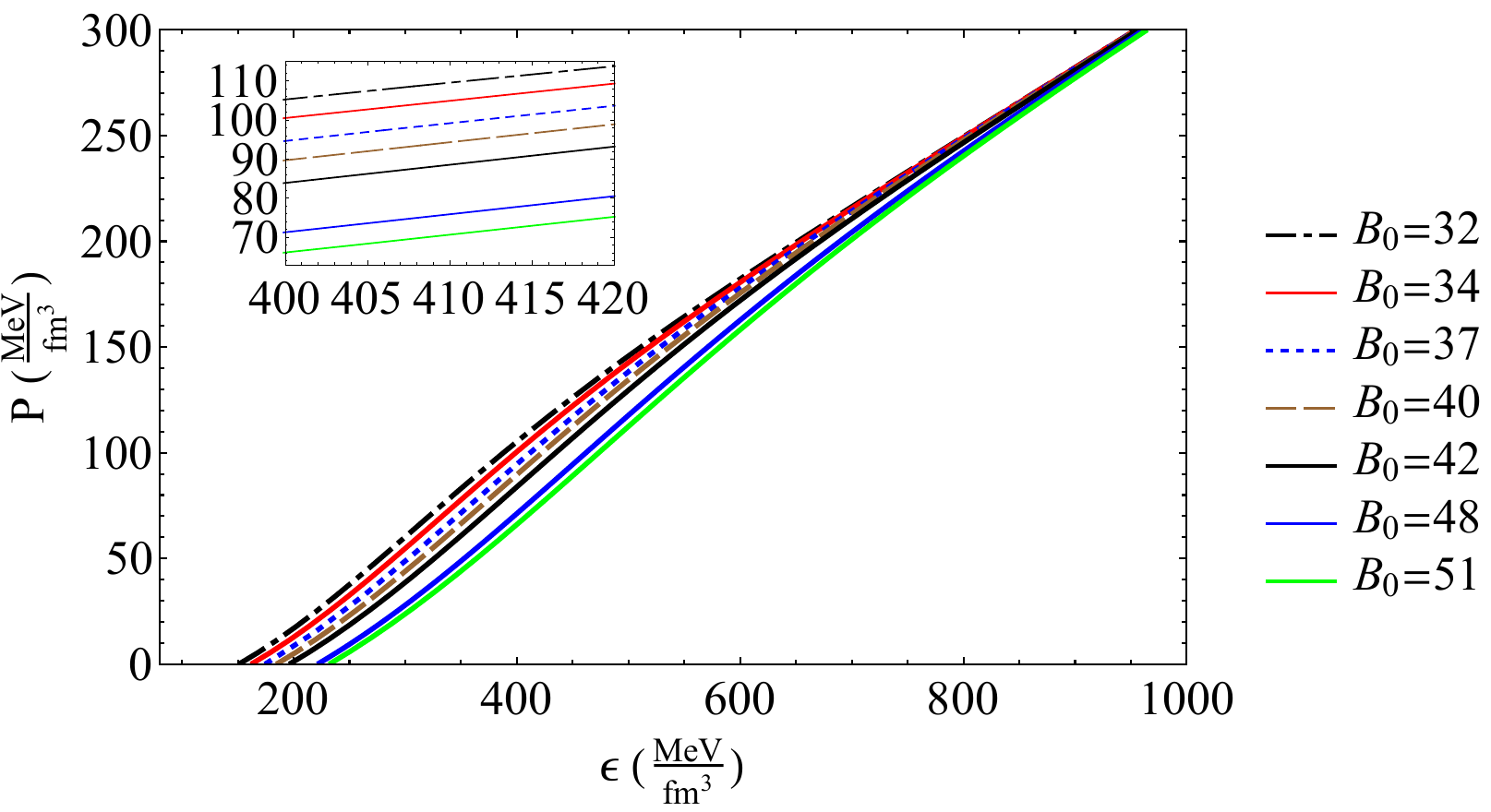}
		{a=0.4}
	\end{subfigure}
	\hfill
	\begin{subfigure}{0.45\textwidth}
		\centering
		\includegraphics[width=\textwidth]{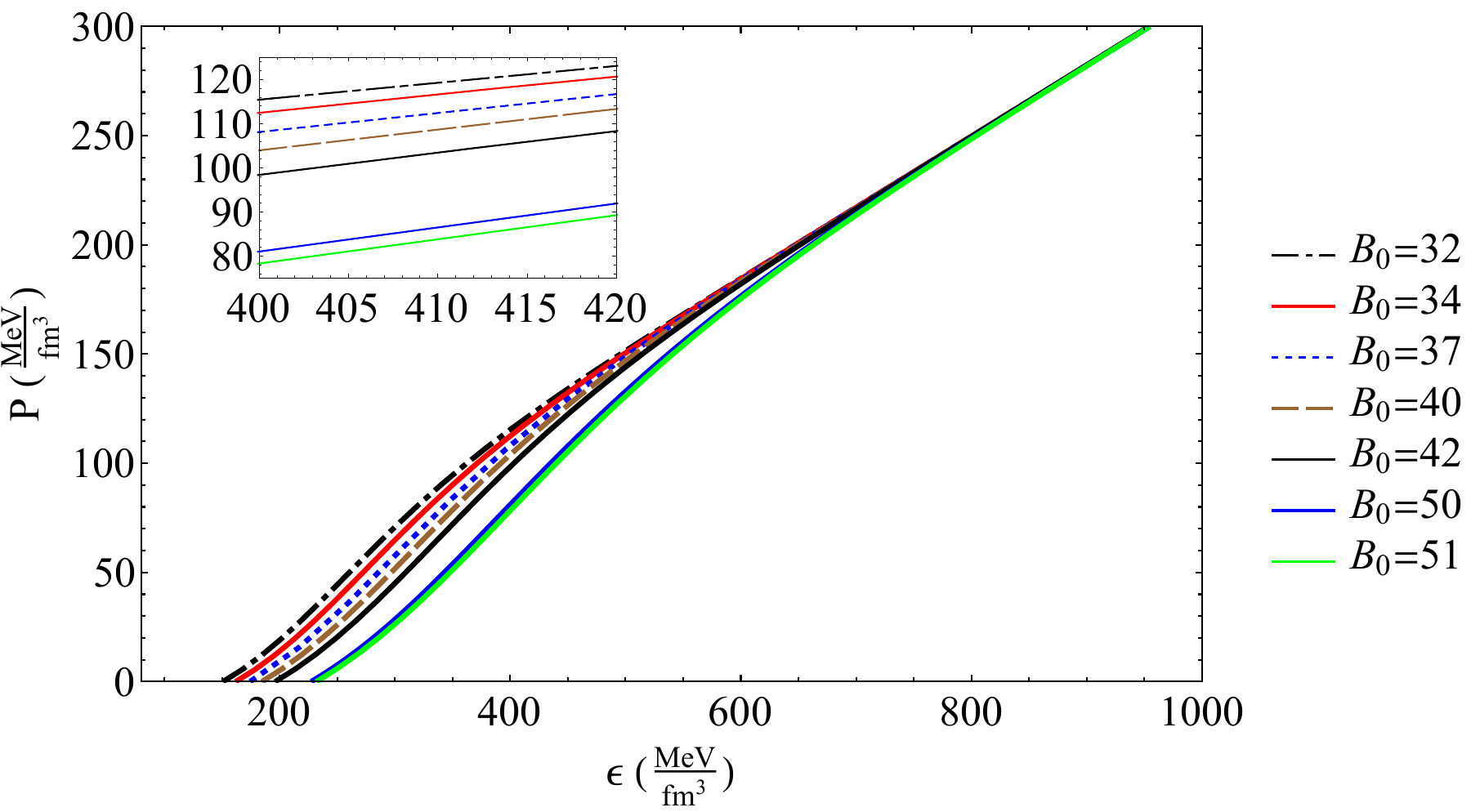}
		{a=0.8}
	\end{subfigure}
	\caption{EOS for different values of $a$ and $B_0$.}
	\label{different EOSs}
\end{figure}

\begin{figure}[h!]
	\centering
	\begin{subfigure}{0.45\textwidth}
		\centering
		\includegraphics[width=\textwidth]{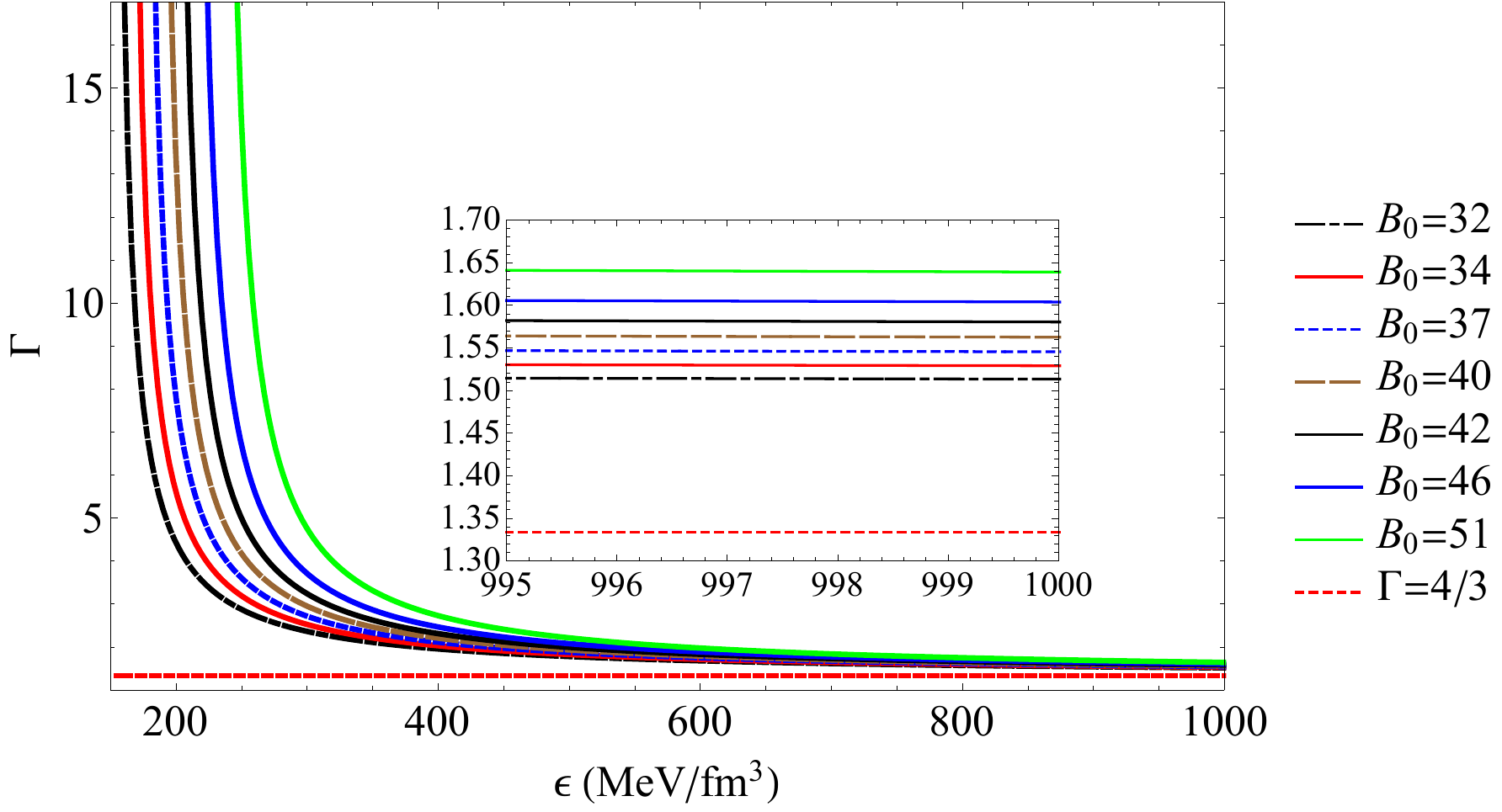}
		{a=0.1}
	\end{subfigure}
	\hfill
	\begin{subfigure}{0.45\textwidth}
		\centering
		\includegraphics[width=\textwidth]{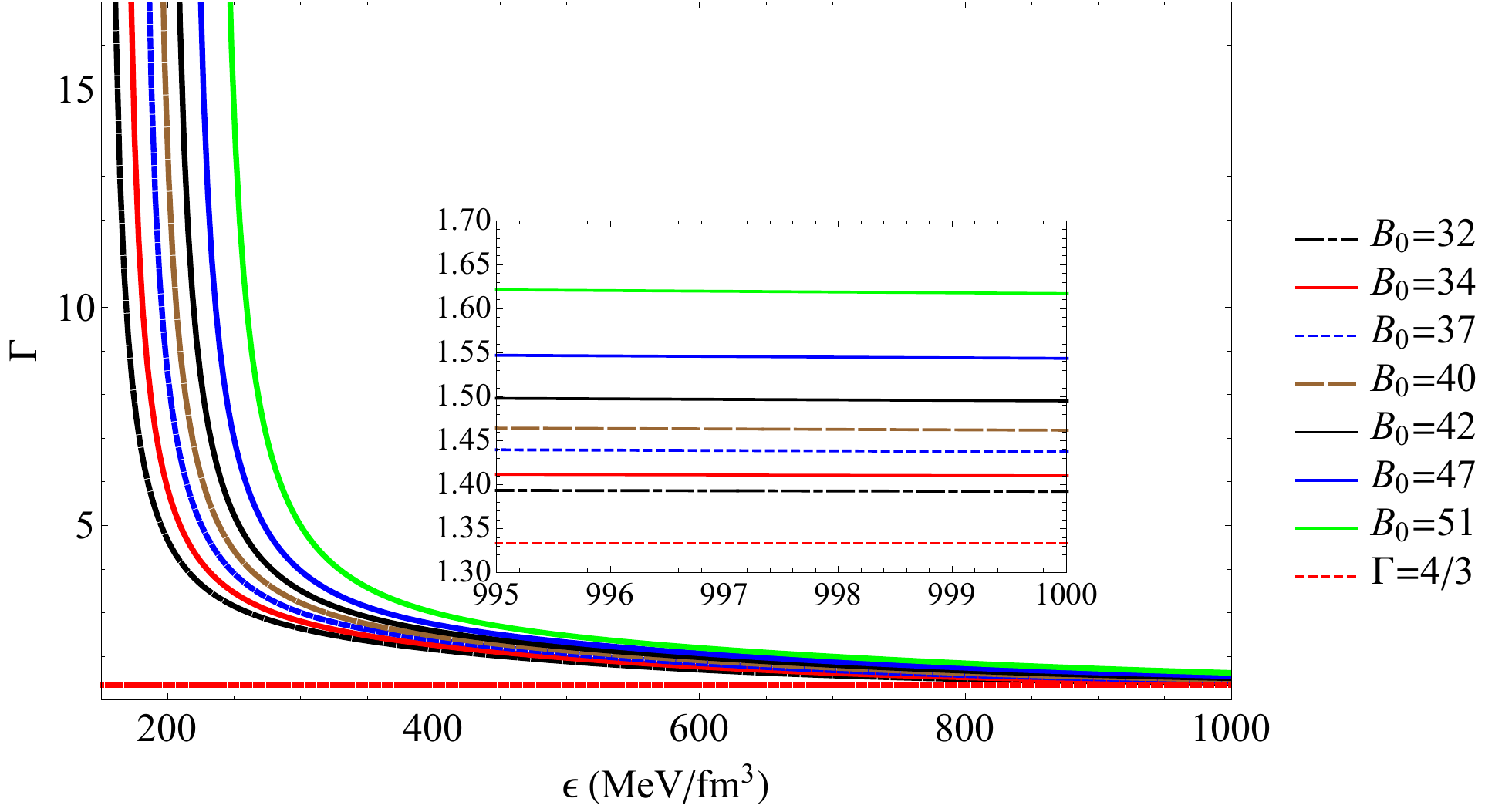}
		{a=0.2}
	\end{subfigure}
	\hfill
	\begin{subfigure}{0.45\textwidth}
		\centering
		\includegraphics[width=\textwidth]{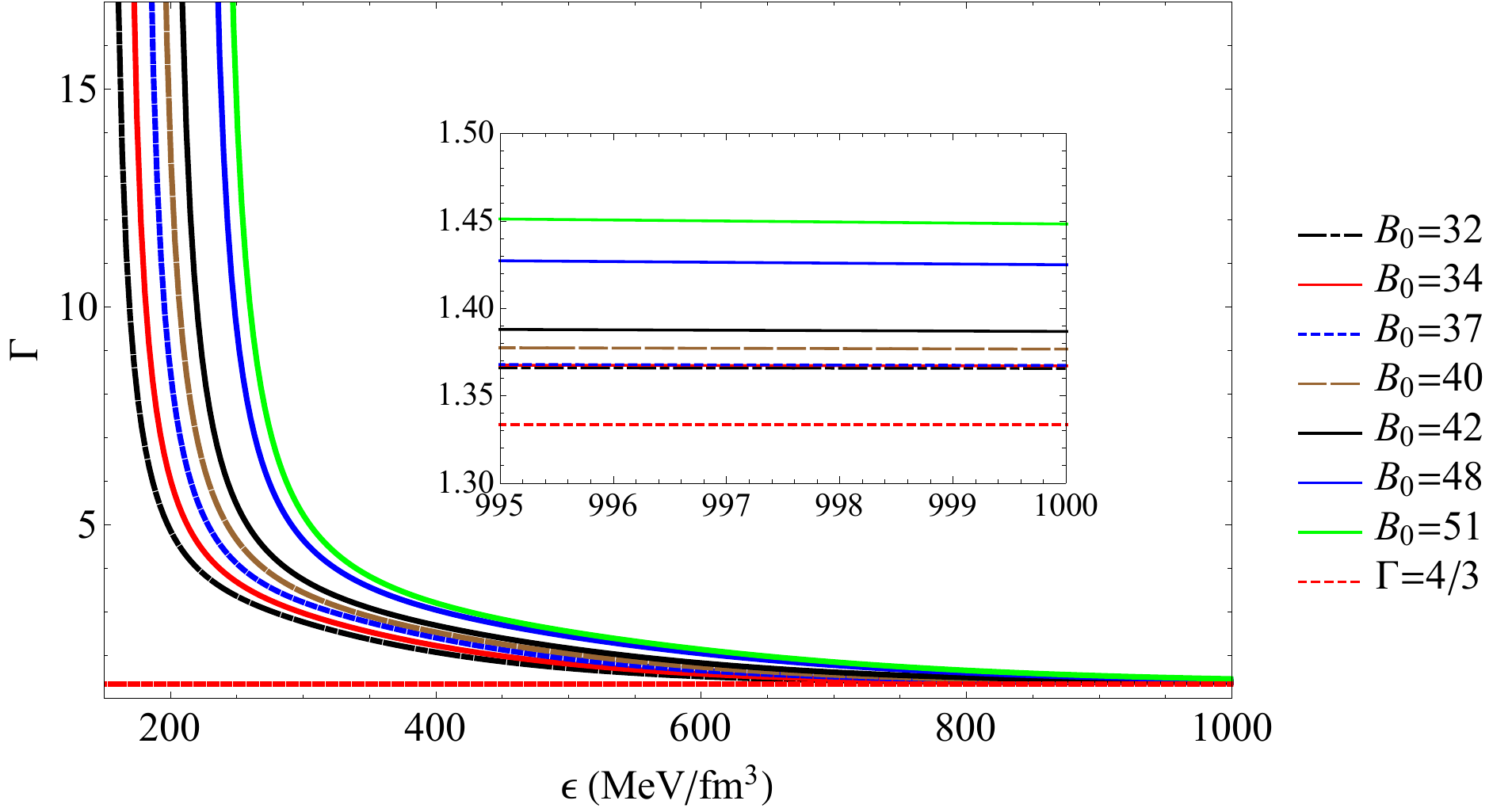}
		{a=0.4}
	\end{subfigure}
	\hfill
	\begin{subfigure}{0.45\textwidth}
		\centering
		\includegraphics[width=\textwidth]{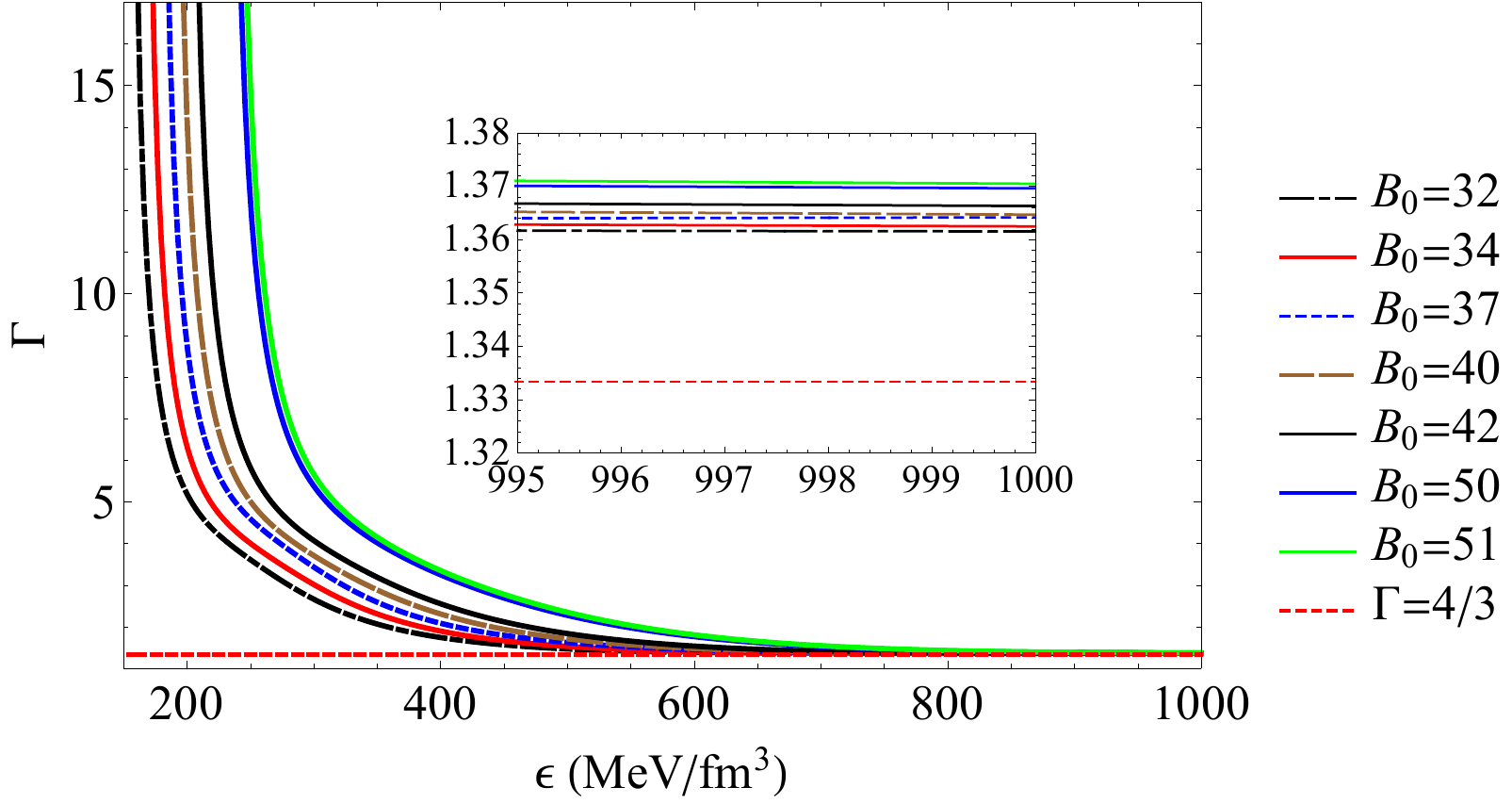}
		{a=0.8}
	\end{subfigure}
	\caption{Adiabatic index for different values of $a$ and $B_0$.}
	\label{Adias}
\end{figure}

\begin{figure}[h!]
	\centering
	\begin{subfigure}{0.45\textwidth}
		\centering
		\includegraphics[width=\textwidth]{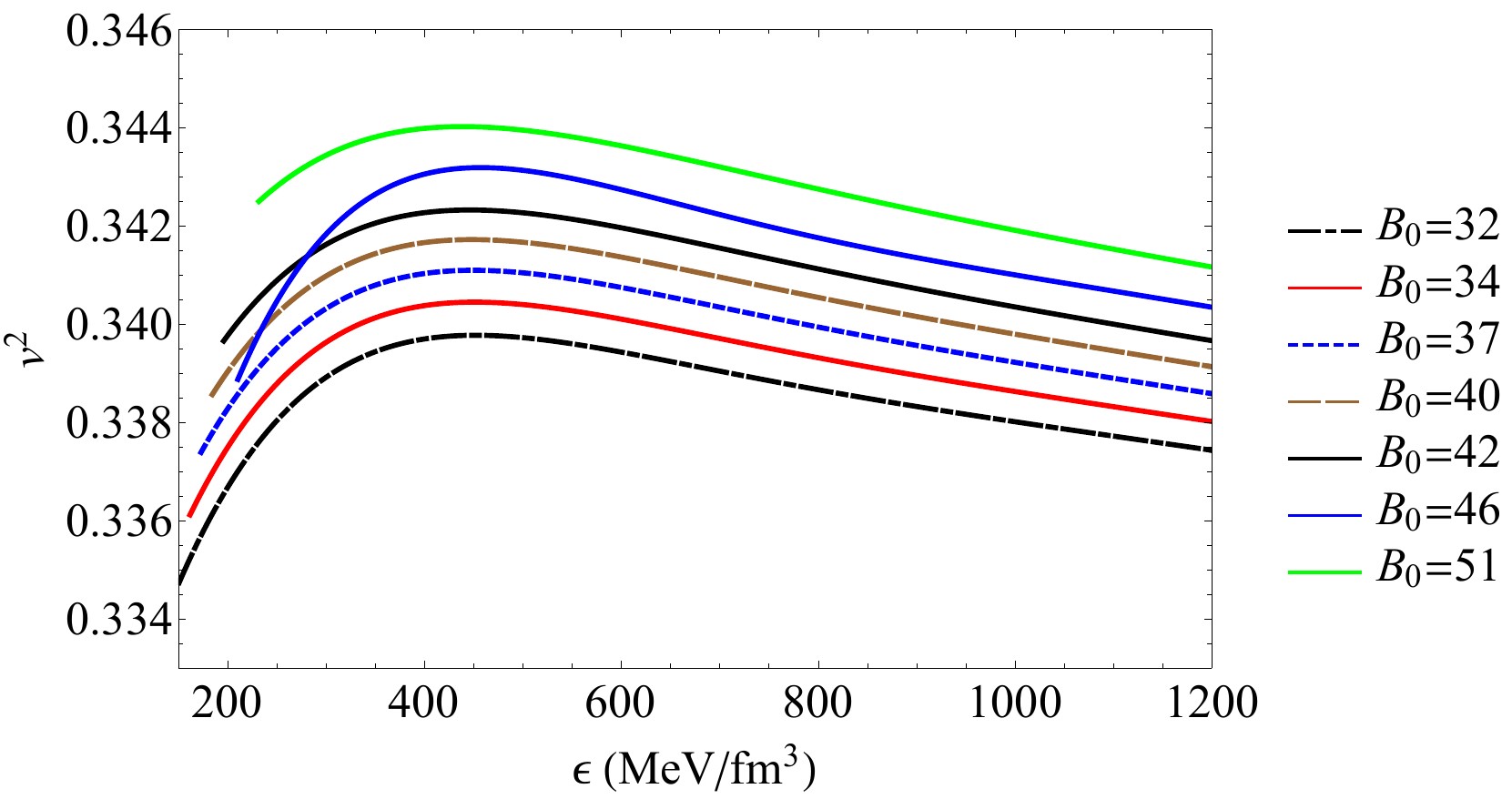}
{a=0.1}
	\end{subfigure}
	\hfill
	\begin{subfigure}{0.45\textwidth}
		\centering
		\includegraphics[width=\textwidth]{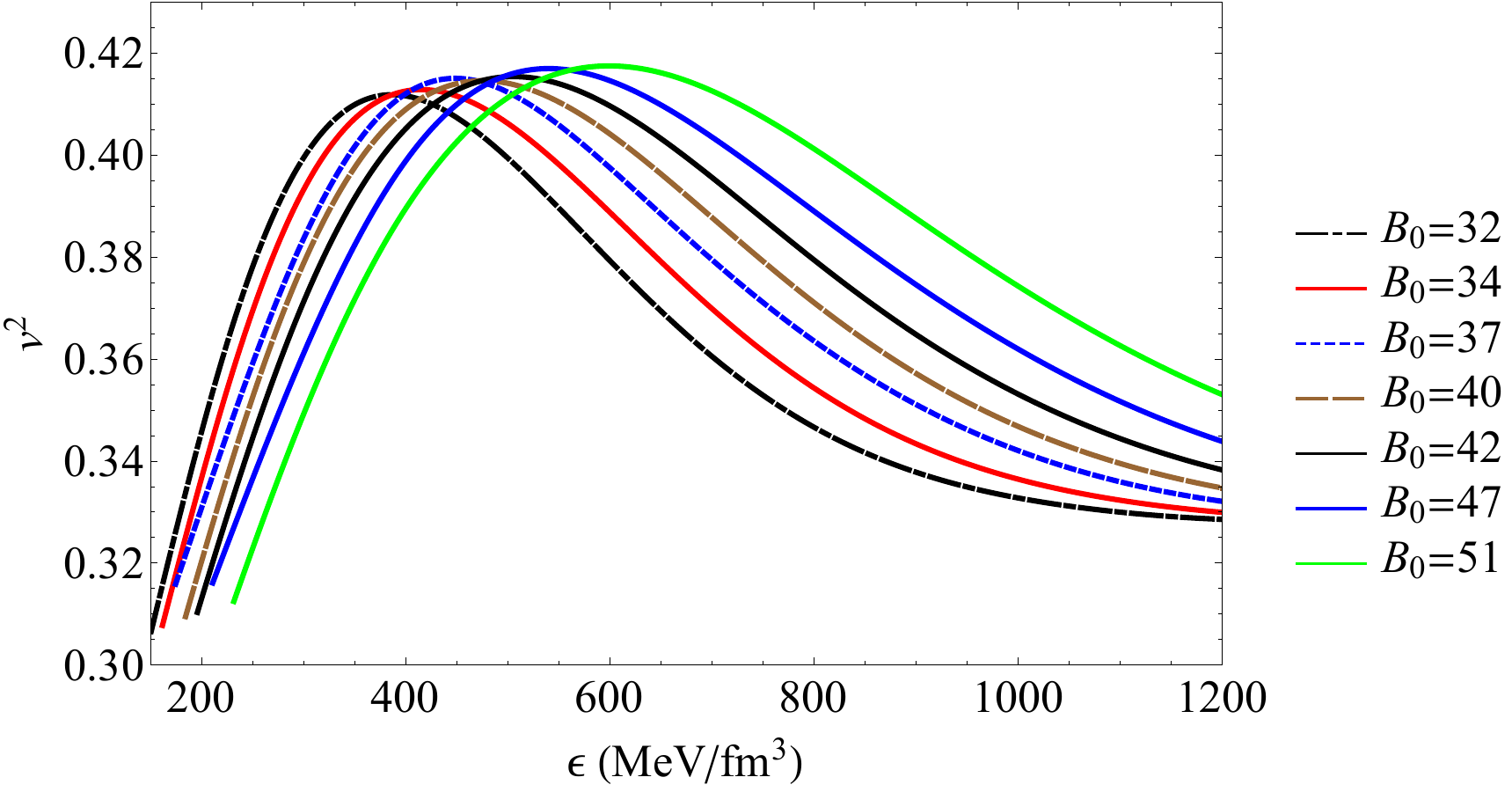}
{a=0.2}
	\end{subfigure}
	\hfill
	\begin{subfigure}{0.45\textwidth}
		\centering
		\includegraphics[width=\textwidth]{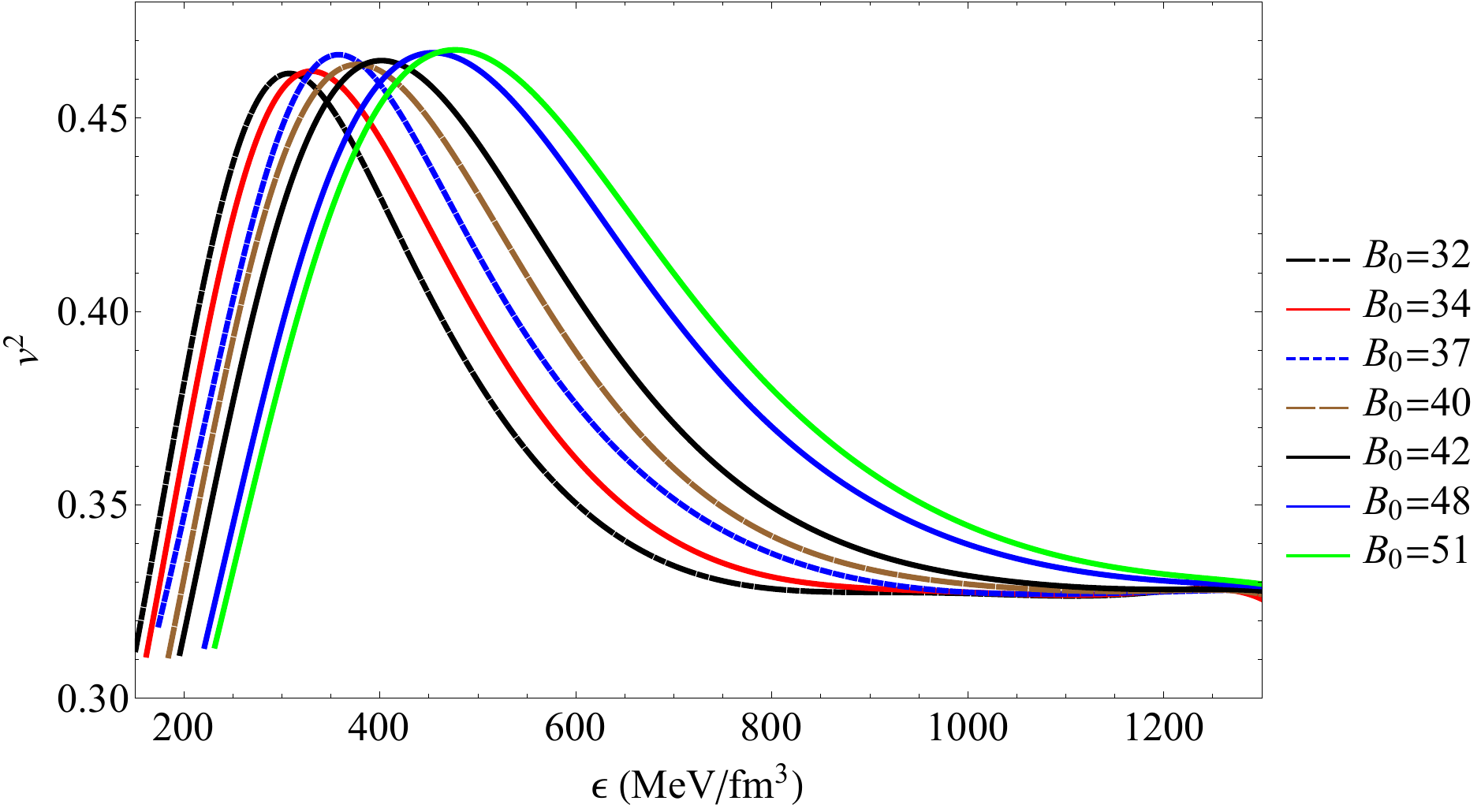}
	{a=0.4}
	\end{subfigure}
	\hfill
	\begin{subfigure}{0.45\textwidth}
		\centering
		\includegraphics[width=\textwidth]{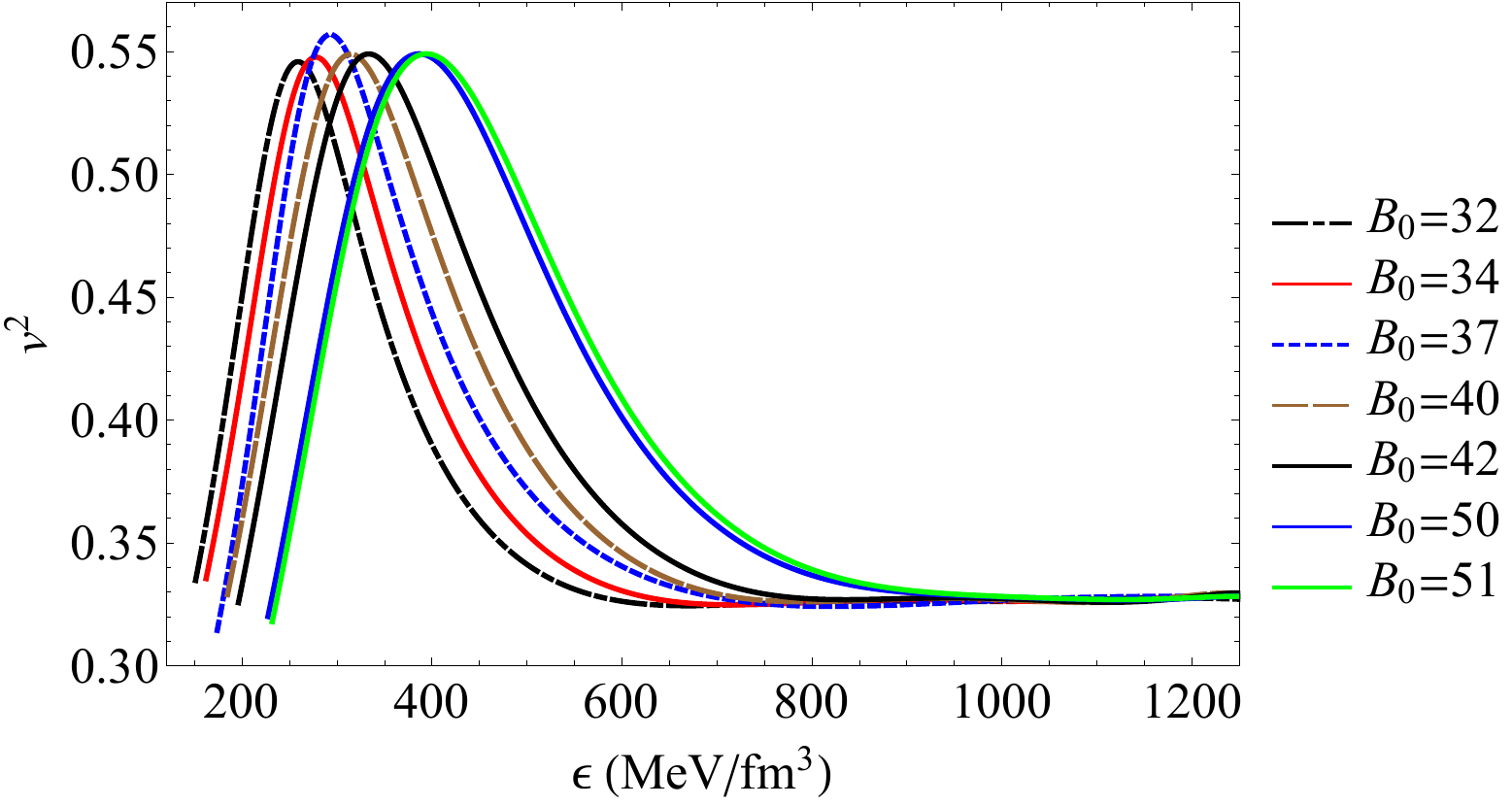}
	{a=0.8}
	\end{subfigure}
	\caption{Speed of sound (in the unit of light speed $c$) for different values of $a$ and $B_0$.}
\label{different sound speeds}
\end{figure}

In this section, we derive the thermodynamic properties of SQM, including the EOS, adiabatic index, and speed of sound, considering a density-dependent $B$. We first obtain the EOS for various values of $a$ and $B_0$. We then examine how the EOS changes as the value of $a$ increases. Fig. \ref{different EOSs} presents various EOSs of SQM for different values of $a$ and $B_0$. As shown in Fig. \ref{different EOSs}, increasing $a$ causes the behavior of the EOSs to deviate from linearity. This feature is particularly evident when comparing the EOSs with $a=0.1$ and $a=0.8$. Now, we need to check whether the new EOSs meet the conditions for dynamical stability and causality. Specifically, we will examine how the new EOSs affect the behavior of the  speed of sound.  
Fig. \ref{Adias} presents the adiabatic index for different values of $a$ and $B_0$. The figure shows that the constraint $\Gamma > \frac{4}{3}$ is met for all EOSs, indicating that dynamical stability is well established.
As we know, the behavior of the speed of sound directly impacts the stiffness of the EOS. It plays a crucial role in determining both the maximum mass ($M_{\text{TOV}}$) of compact stars and their tidal deformability parameter ($\Lambda$). Fig. \ref{different sound speeds} shows the speed of sound versus energy density for different values of $a$ and $B_0$. 
The figure illustrates not only well-satisfied causality but also the graphs' convergence toward the value  $v^2/c^2=\frac{1}{3}$ at high energy densities. This behavior is essential because at high densities, the speed of sound is anticipated to approach asymptotically that of a relativistic gas without interaction. Furthermore, upon comparing Fig. \ref{different sound speeds} with Fig. \ref{sound1}, it becomes evident that the speed of sound increases when considering a  density-dependent $B$. Fig. \ref{different sound speeds} demonstrates that as the value of $a$ increases, the speed of sound increases as well. This characteristic can contribute to an increase in $M_{TOV}$. However, while larger masses may be obtained, it is imperative to ensure that the $\Lambda$ condition also holds. Therefore, it is necessary to re-examine the $\Lambda$ behavior for the new EOSs. Our investigation showed that the $\Lambda$ condition is satisfied only for $a\lesssim 0.8$. Hence, in the following, both the $M-R$ and $\Lambda-M$ diagrams are investigated for these values of $a$. 
\section{Structral properties of SQS for different values of $a$ and $B_0$}
In this section, we explore different values of $a$ and $B_0$ in \ref{DDbag} and investigate their effects on the structural properties of SQS. Specifically, we aim to identify values of $a$ and $B_0$ that result in EOSs of SQM, yielding a mass for a pure SQS that exceeds $2M_{\odot}$. Simultaneously, these values must satisfy observational constraints, such as tidal deformability.
\subsection{Mass and radius}
\begin{figure}[h!]
	\centering
	\begin{subfigure}{0.45\textwidth}
		\centering
		\includegraphics[width=\textwidth]{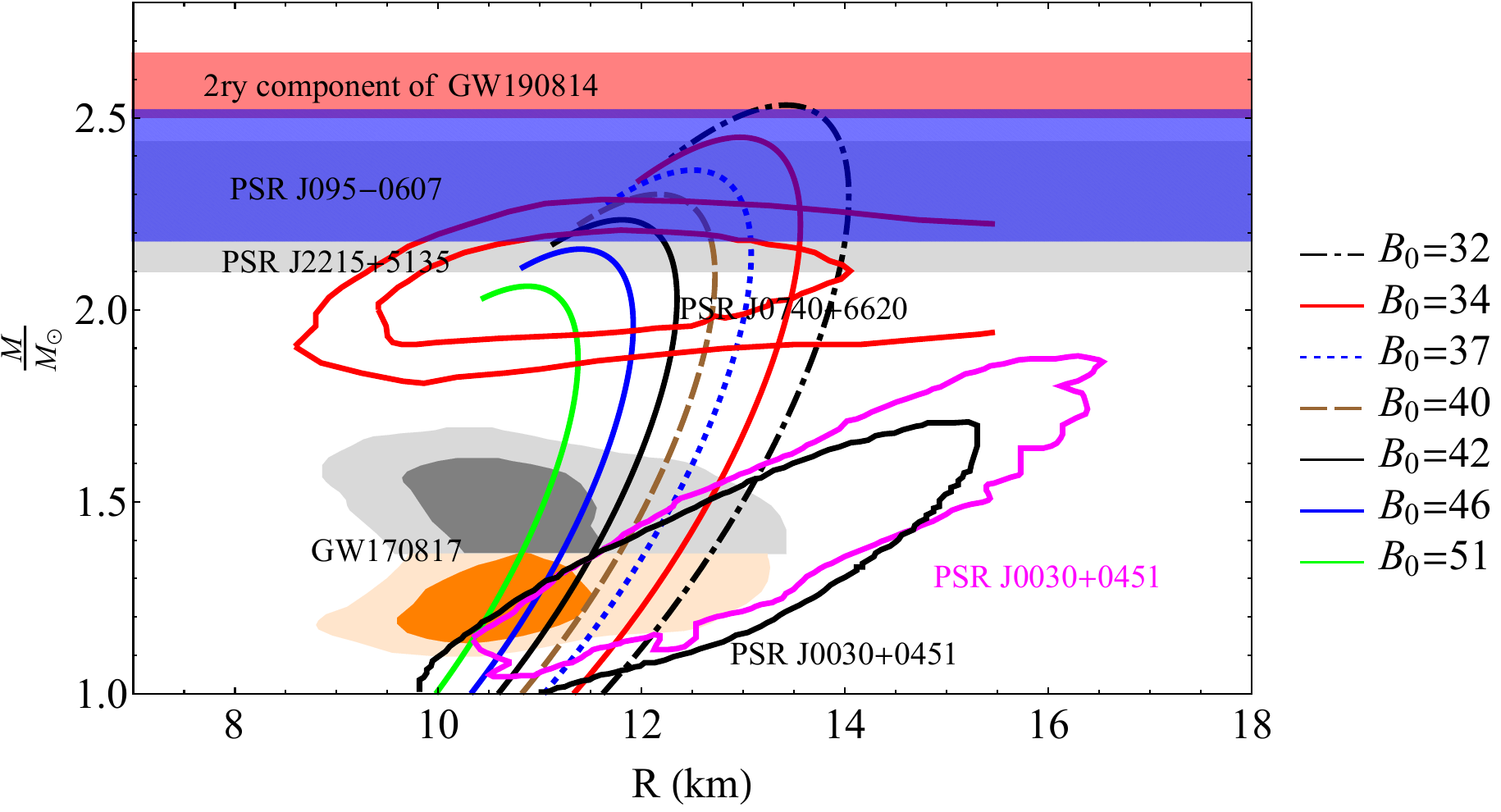}
{a=0.1}
	\end{subfigure}
	\hfill
	\begin{subfigure}{0.45\textwidth}
		\centering
		\includegraphics[width=\textwidth]{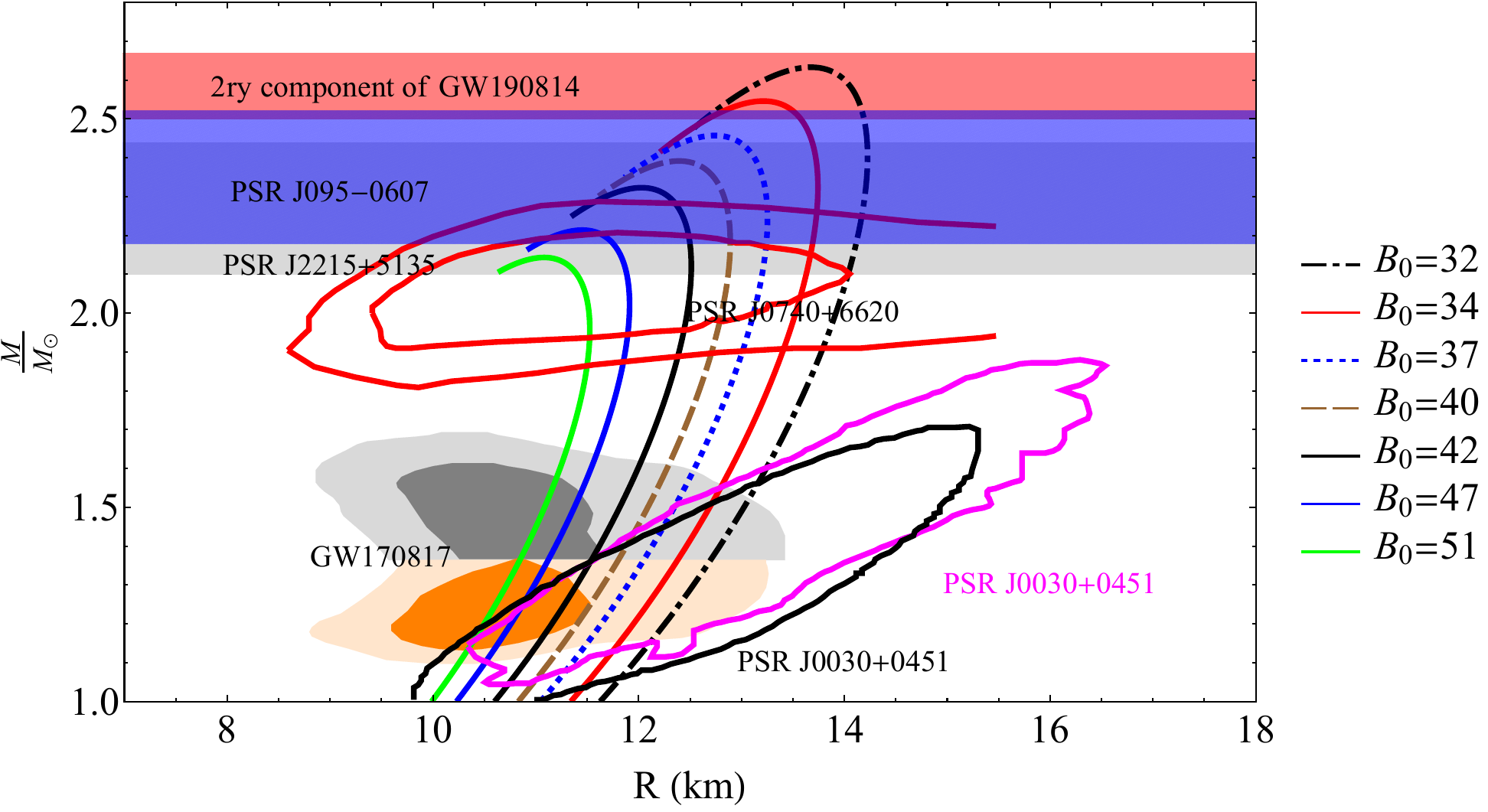}
{a=0.2}
	\end{subfigure}
	\hfill
	\begin{subfigure}{0.45\textwidth}
		\centering
		\includegraphics[width=\textwidth]{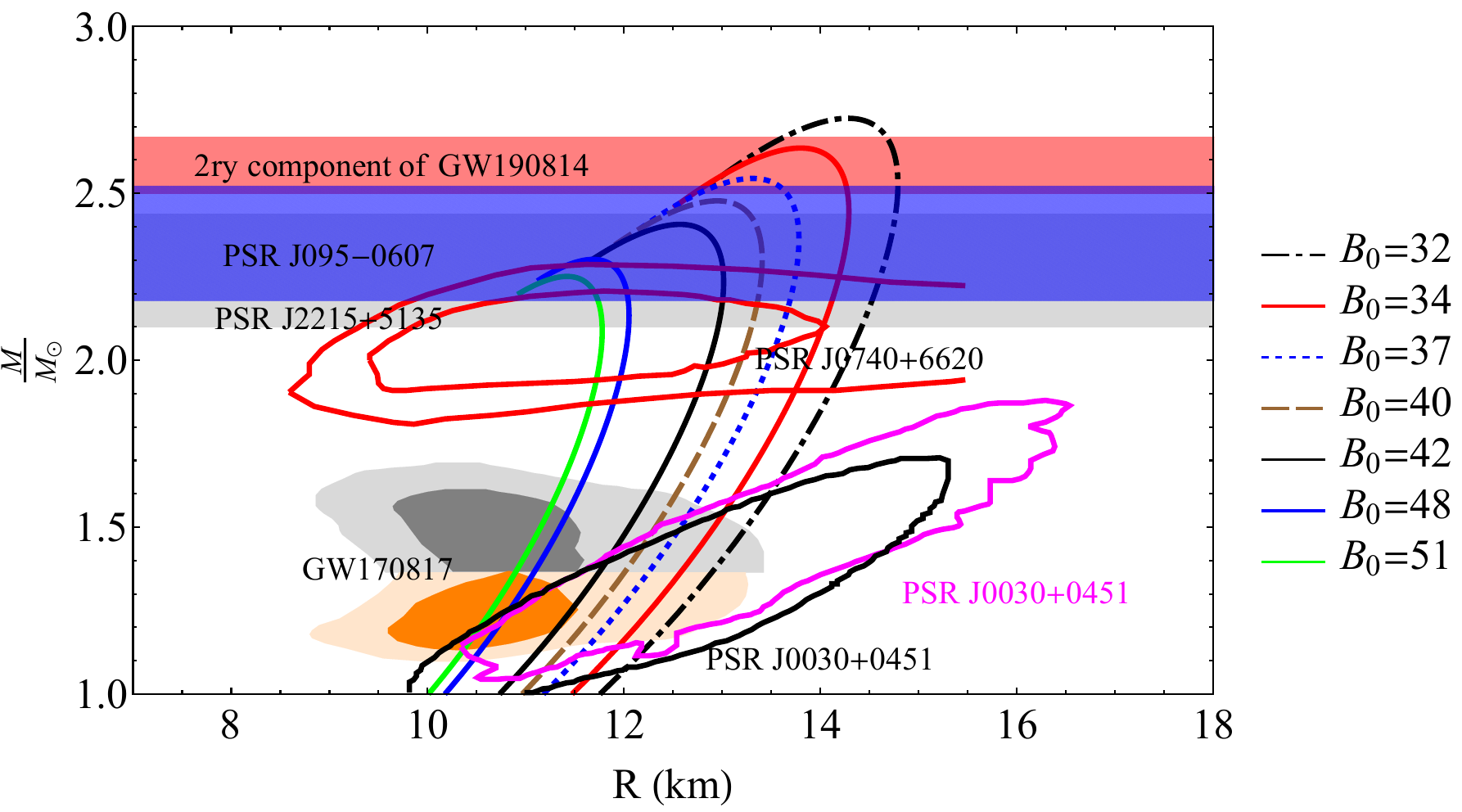}
		{a=0.4}
	\end{subfigure}
	\hfill
	\begin{subfigure}{0.45\textwidth}
		\centering
		\includegraphics[width=\textwidth]{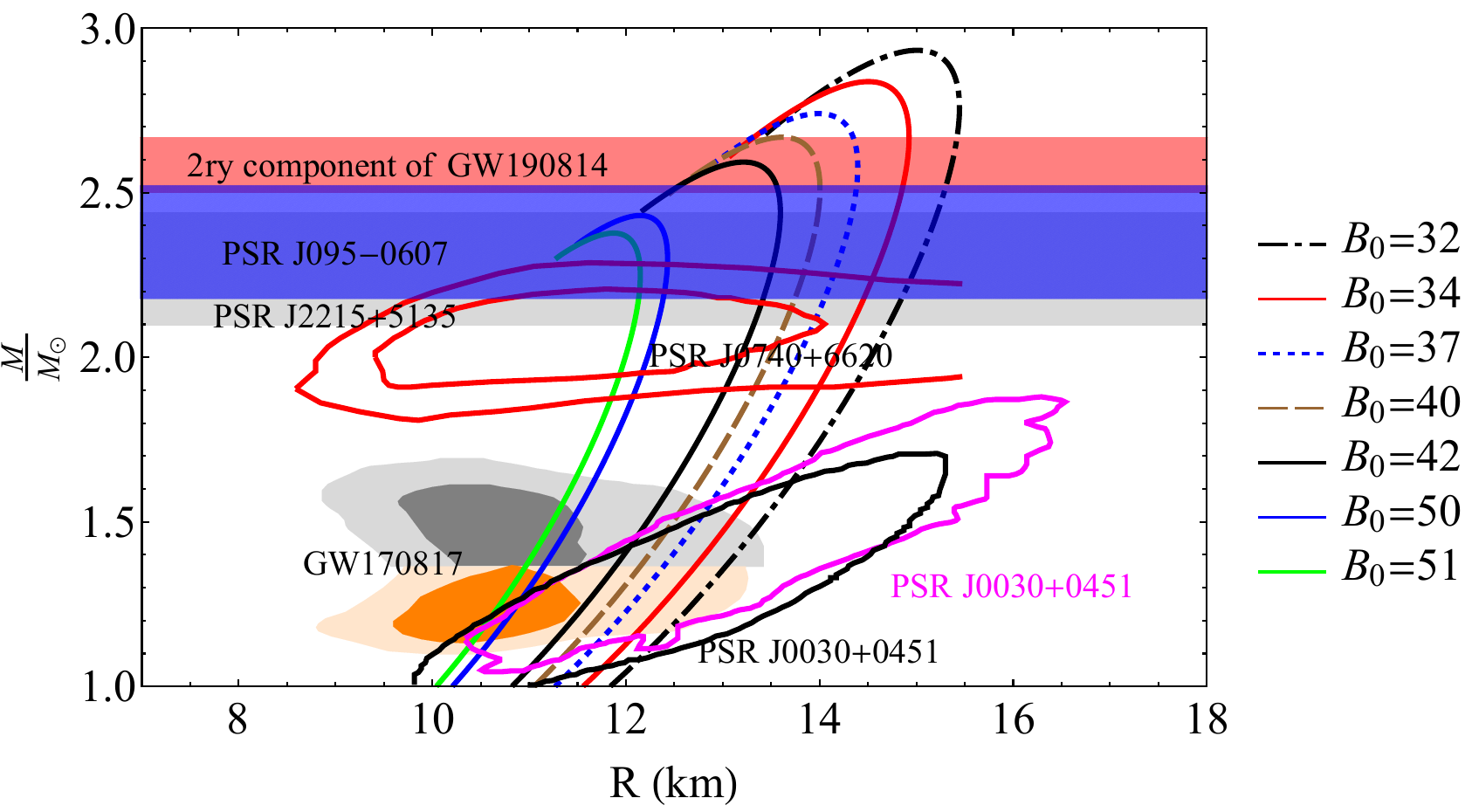}
	{a=0.8}
	\end{subfigure}
	\caption{$M-R$ diagram for different values of $a$ and $B_0$.}
	\label{different m-r diagrams}
\end{figure}
Fig. \ref{different m-r diagrams} represents $M-R$ diagram for different values of $a$ and various values of $B_0$ that meet the stability condition. When comparing this figure to Fig. \ref{different m-r diagrams}, it becomes evident that the inclusion of a  density-dependent $B$ leads to an increase in $M_{TOV}$. As mentioned earlier, increasing the speed of sound by raising the value of $a$ should also increase the $M_{TOV}$. Fig. \ref{different m-r diagrams}. This characteristic is evident in Fig. \ref{different m-r diagrams} ; as the value of $a$ increases from $0.1$ to $0.8$, the mass of the star also increases. Additionally, the results remain consistent with the permitted regions of mass and radius observed. For values $a\gtrsim0.2$, the results encompass PSR J2215+5135, PSR J095+0607, and even the secondary component of GW190814 quite effectively. However, it is crucial to verify the $\Lambda$ constraint. In the following section, we will demonstrate that when the value of 
$a$ is held constant, the $\Lambda$ condition is not satisfied for all values of $B_0$. Consequently, although Fig. \ref{different m-r diagrams} shows that the obtained SQSs with masses exceeding $2M_{\odot}$ meet the conditions related to mass and radius, not all of them satisfy the $\Lambda$ condition.
\subsection{Tidal deformability}
So far, our results have closely followed the mass and radius constraints derived from various pulsars and binary gravitational waves (see Fig. \ref{different m-r diagrams}). In this section, we investigate the behavior of dimensionless tidal deformability versus the mass of SQS for different values of $a$ and $B_0$. Although our primary focus for $\Lambda$ is $70 < \Lambda_{1.4\textup{M}_\odot} < 580$, derived from GW170817, we have also analyzed whether the results meet the constraint on $\Lambda$ from GW190814. If we assume the secondary mass of GW190814 is that of a neutron star, then the parameter $\Lambda_{1.4M_\odot}$ is constrained to $458 < \Lambda_{1.4M_\odot} < 889$ \cite{Abbott2020}. However, the former constraint is more widely accepted in the literature. Therefore, we divide this section into two parts, addressing the $\Lambda$ constraints from GW170817 and GW190814, respectively. 
\subsubsection{Testing the results with the $\Lambda$ constraint from GW170817.}
\begin{figure}[h!]
		\centering
	\begin{subfigure}{0.45\textwidth}
		\centering
		\includegraphics[width=\textwidth]{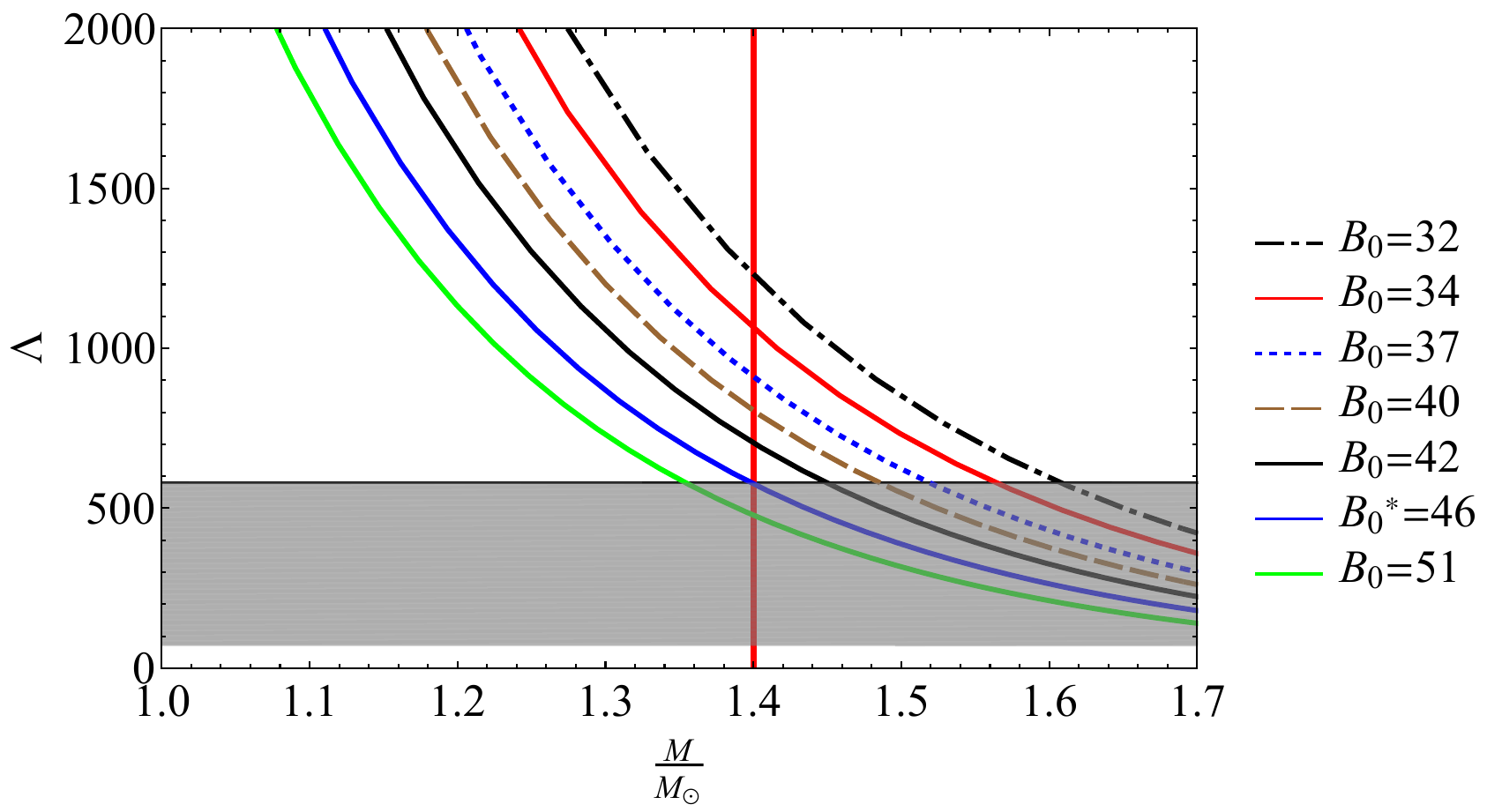}
	{a=0.1}
	\end{subfigure}
	\hfill
	\begin{subfigure}{0.45\textwidth}
		\centering
		\includegraphics[width=\textwidth]{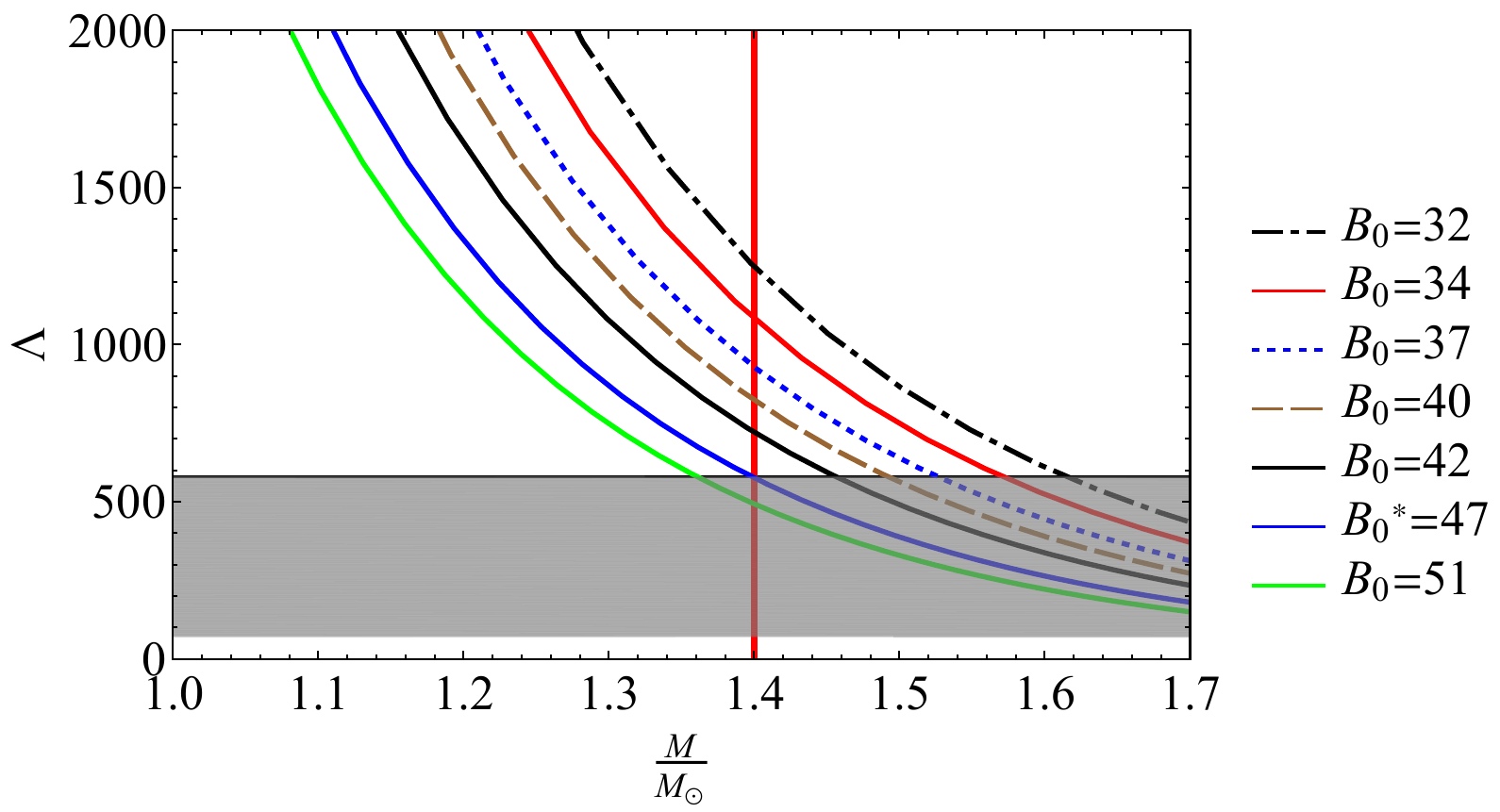}
		{a=0.2}
	\end{subfigure}
	\hfill
	\begin{subfigure}{0.45\textwidth}
		\centering
		\includegraphics[width=\textwidth]{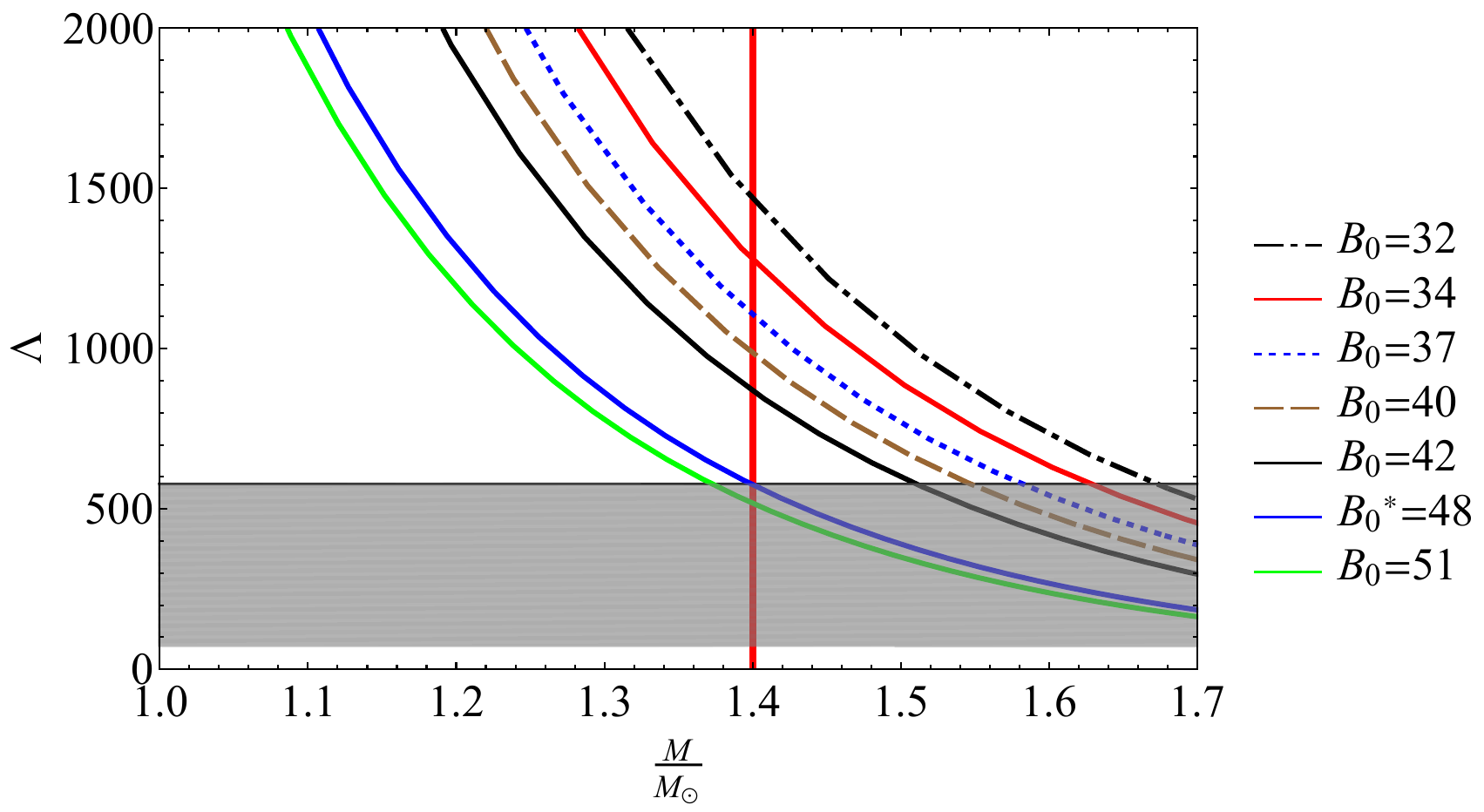}
	{a=0.4}
	\end{subfigure}
	\hfill
	\begin{subfigure}{0.45\textwidth}
		\centering
		\includegraphics[width=\textwidth]{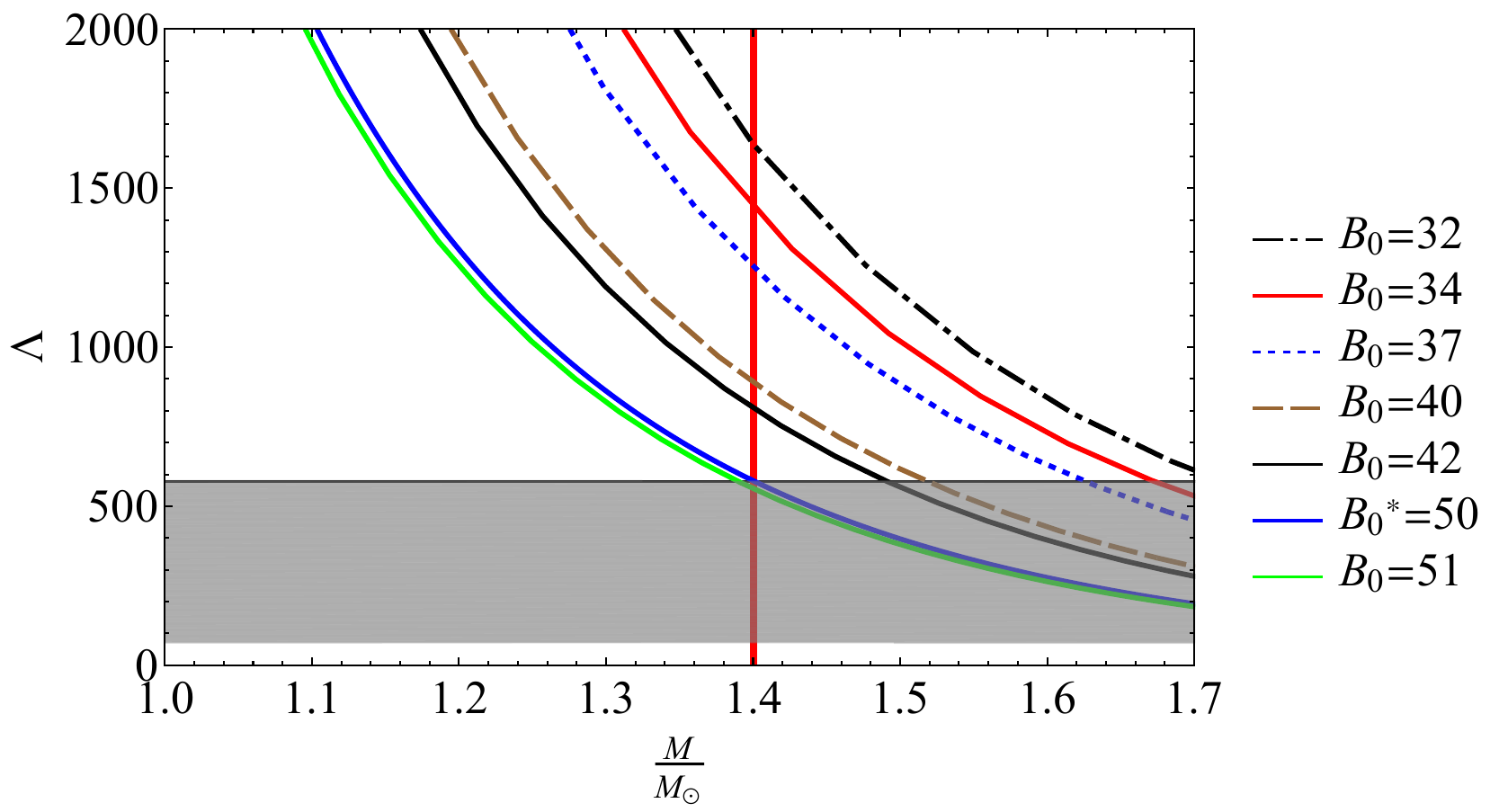}
		{a=0.8}
	\end{subfigure}
	\caption{$\Lambda-M$ diagram for different values of $a$ and $B_0$. The Gray area shows the constraint $70 \lesssim \Lambda \lesssim 580$ from GW170817.}
	\label{different tidals}
\end{figure}

Fig. \ref{different tidals} shows $\Lambda-M$ diagram for different values of $a$ and $B_0$. This figure indicates that the range of results satisfying the constraint $70 \lesssim \Lambda \lesssim 580$ (represented by the gray bar in Fig. \ref{different tidals}) becomes narrower as the value of $a$ increases. In Fig. \ref{different tidals}, for each value of $a$, there is a specific value of $B_0$ denoted by ${B_0}^{*}$. ${B_0}^{*}$ represents the value of $B_0$ for which the corresponding EOS yields the maximum possible mass of SQS while satisfying the constraint $70 \lesssim \Lambda \lesssim 580$. Table \ref{results2} represents the structural properties of SQS for different values of $a$ and $B_0$. According to Table \ref{results2}, for $a=0.1$, the only allowed values of $B_0$ that satisfy observational constraints are in the range $46\lesssim B_0 \lesssim 51$. Upon comparing the results of Tables \ref{results1} and \ref{results2}, it is evident that the maximum allowed mass of SQS extends from $2.03\textup{M}_\odot$ to $2.16\textup{M}_\odot$, representing a significant increase. This augmentation in mass holds importance as the updated value still adheres well to the observational constraints. The results in Table \ref{results2} show that as $a$ increases, a smaller range of $B_0$ values can yield EOSs that satisfy the constraint  $\Lambda_{1.4\textup{M}_\odot} < 580$. 
For $a=0.2$, the maximum allowed $M_{TOV}$ is $2.21\textup{M}_\odot$, which accurately describes PSR J0740+6620.  When $a$ increases to 0.4, $M_{TOV}$ reaches $2.30\textup{M}_\odot$, encompassing the masses of PSR J2215+5135 and PSR J095-0607. Finally, for $a=0.8$, $M_{TOV}$ can reach $2.43\textup{M}_\odot$ and $2.38\textup{M}_\odot$ for $B_0=50 \text{ MeV/fm}^3$ and $B_0=52 \text{ MeV/fm}^3$, respectively, both of which accurately describe PSR J2215+5135 and PSR J095-0607.
 
\begin{table}[h!]
	\caption{		
		Structural properties of SQS for different values of $a$ and $B_0$, comparing with the $\Lambda$ constraint from GW170817.}
	\centering
	\small
	\begin{subtable}{1.0\linewidth}
		\centering
		\scalebox{0.77}{
			\begin{tabular}{|c|c|c|c|c|c|c|c|}
				\hline
				\multicolumn{5}{|c|}{\textbf{a=0.1}} \\
				\hline
				$B_0(MeV/fm^3)$ & $\Lambda_{1.4\textup{M}_\odot}$& $R(km)$ & $M_{TOV}(\textup{M}_\odot)$ & GW170817 \\	\hline
				$32$ & 1237.10 & 13.41 & 2.53 & $\times$ \\ \hline
				$34$ & 1051.20 & 12.96 & 2.45 & $\times$  \\ \hline
				$37$ & 911.72 & 12.50 & 2.36 & $\times$  \\ \hline
				$40$ & 800.32 & 12.15 & 2.30 & $\times$  \\ \hline
				$42$ & 704.62 & 11.80 & 2.23 & $\times$  \\ \hline
				\textcolor{black}{$B_0^*=46$}& \textcolor{black}{579.69} & \textcolor{black}{11.33} & \textcolor{black}{2.16} & \checkmark  \\ \hline
				$51$ & 477.13 & 10.88 & 2.06 & \checkmark  \\ \hline
			\end{tabular}
}
	
	\end{subtable}%
\\
	\begin{subtable}{1.0\linewidth}
		\centering
		\scalebox{0.77}{
			\begin{tabular}{|c|c|c|c|c|c|c|}
					\hline
				\multicolumn{5}{|c|}{\textbf{a=0.2}} \\
				\hline
				$B_0(MeV/fm^3)$ & $\Lambda_{1.4\textup{M}_\odot}$& $R(km)$ & $M_{TOV}(\textup{M}_\odot)$ & GW170817\\\hline
				$32$ & 1240.59 & 13.68 & 2.63 & $\times$ \\ \hline
				$34$ & 1079.20 & 13.21 & 2.54 & $\times$ \\ \hline
				$37$ & 921.98 & 12.74 & 2.46 & $\times$ \\ \hline
				$40$ & 825.85 & 12.39 & 2.39 & $\times$\\ \hline
				$42$ & 718.19 & 12.03 & 2.32 & $\times$\\ \hline
				\textcolor{black}{$B_0^*=47$}& \textcolor{black}{576.28} & \textcolor{black}{11.44} & \textcolor{black}{2.21} & \checkmark\\ \hline
				$51$ & 494.84 & 11.08 & 2.14  &\checkmark\\ \hline
		\end{tabular}}
		
	\end{subtable}
	\\
	\begin{subtable}{1.0\linewidth}
		\centering
		\scalebox{0.77}{
			\begin{tabular}{|c|c|c|c|c|c|c|}
					\hline
				\multicolumn{5}{|c|}{\textbf{a=0.4}} \\
				\hline
				$B_0(MeV/fm^3)$ & $\Lambda_{1.4\textup{M}_\odot}$& $R(km)$ & $M_{TOV}(\textup{M}_\odot)$ &GW170817 \\	\hline
				$32$ & 1465.02 & 14.28 & 2.72 & $\times$  \\ \hline
				$34$ & 1273.30 & 13.79 & 2.63 & $\times$\\ \hline
				$37$ & 1102.30 & 13.33 & 2.54 & $\times$ \\ \hline
				$40$ & 986.10 & 12.94 & 2.48 & $\times$ \\ \hline
				$42$ & 867.65 & 12.56 & 2.41 & $\times$\\ \hline
				\textcolor{black}{$B_0^*=48$}& \textcolor{black}{577.23} & \textcolor{black}{11.68} & \textcolor{black}{2.30} & \checkmark \\ \hline
				$51$ & 516.45 & 11.42 & 2.25 & \checkmark\\ \hline
		\end{tabular}}
			\end{subtable}
\\
	\begin{subtable}{1.0\linewidth}
		\centering
		\scalebox{0.77}{
			\begin{tabular}{|c|c|c|c|c|c|c|}
					\hline
				\multicolumn{5}{|c|}{\textbf{a=0.8}} \\
				\hline
				$B_0(MeV/fm^3)$ & $\Lambda_{1.4\textup{M}_\odot}$& $R(km)$ & $M_{TOV}(\textup{M}_\odot)$&GW170817 \\	\hline
				$32$ & 1644.21 & 15.01 & 2.93 & $\times$  \\ \hline
				$34$ & 1441.68 & 14.60 & 2.83 & $\times$ \\ \hline
				$37$ & 1246.71 & 13.98 & 2.74 & $\times$  \\ \hline
				$40$ & 888.34 & 13.61 & 2.66 & $\times$ \\ \hline
				$42$ & 809.02 & 13.21 & 2.59 & $\times$ \\ \hline
				\textcolor{black}{$B_0^*=50$}& \textcolor{black}{577.39} & \textcolor{black}{12.15} & \textcolor{black}{2.43} & \checkmark\\ \hline
				$51$ & 555.84 & 11.86 & 2.38 & \checkmark  \\ \hline
		\end{tabular}}
			\end{subtable}
		\label{results2}
\end{table}	
\subsubsection{Testing the results with the $\Lambda$ constraint from GW190814.}
\begin{figure}[h!]
	\centering
	\begin{subfigure}{0.45\textwidth}
		\centering
		\includegraphics[width=\textwidth]{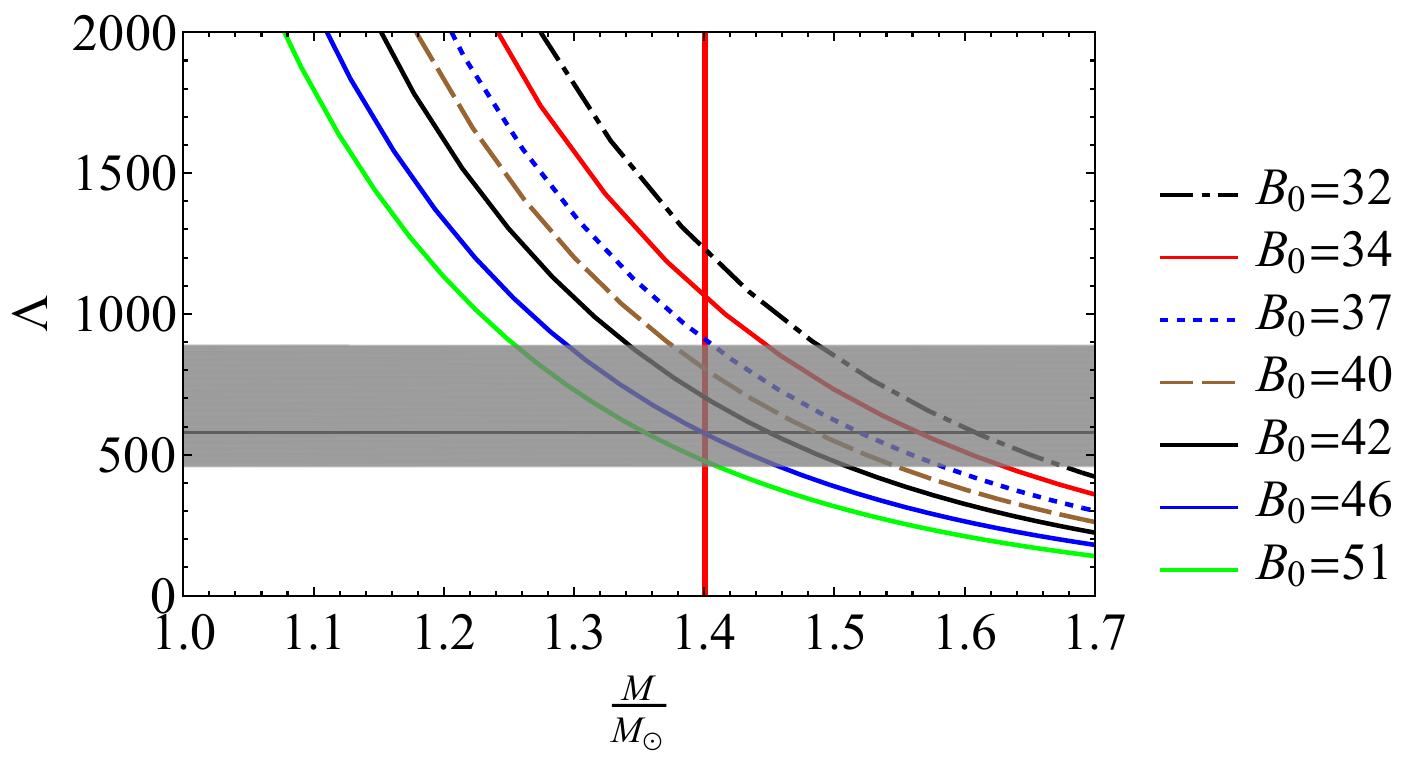}
		{a=0.1}
	\end{subfigure}
	\hfill
	\begin{subfigure}{0.45\textwidth}
		\centering
		\includegraphics[width=\textwidth]{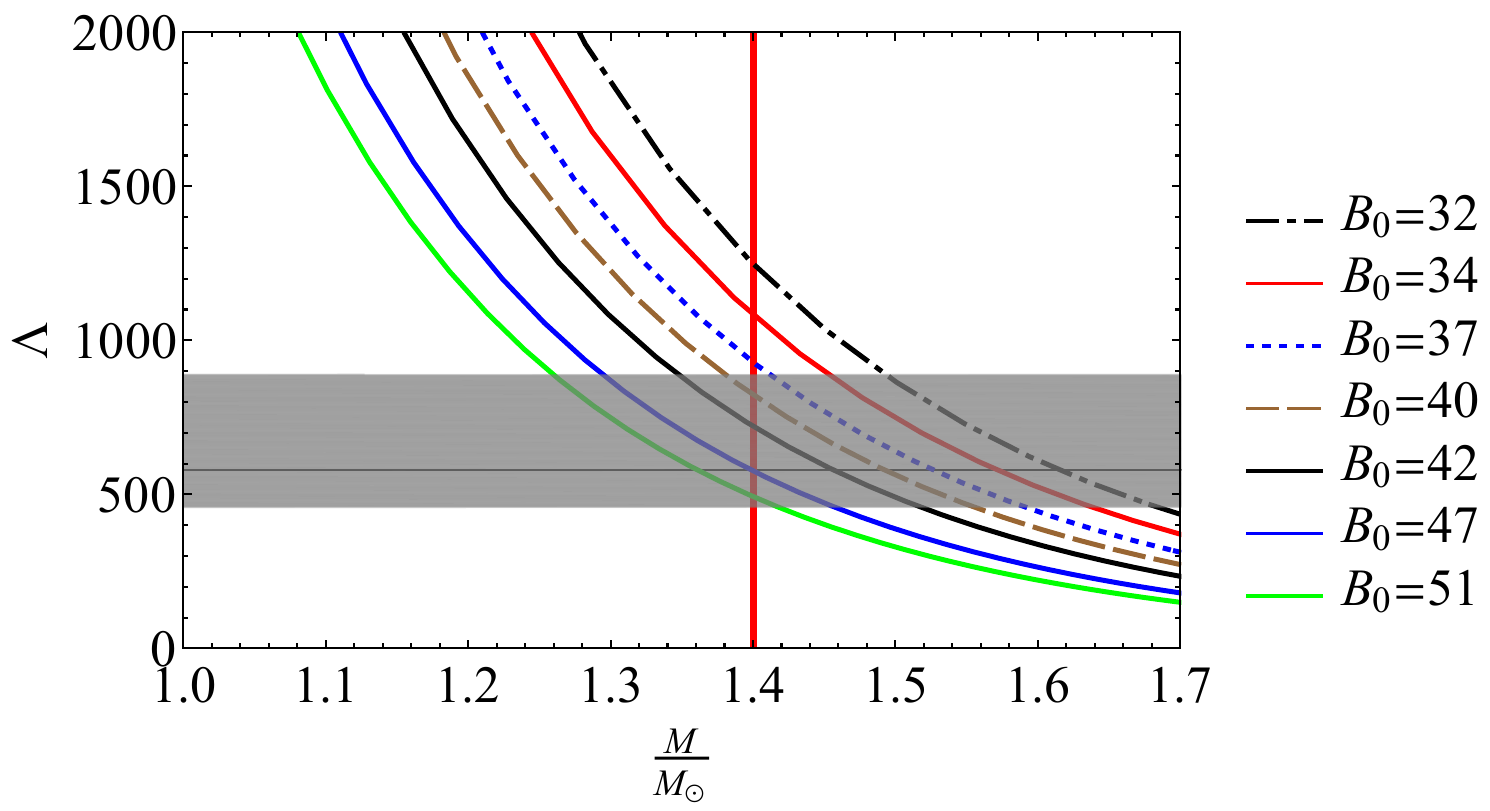}
	{a=0.2}
	\end{subfigure}
	\hfill
	\begin{subfigure}{0.45\textwidth}
		\centering
		\includegraphics[width=\textwidth]{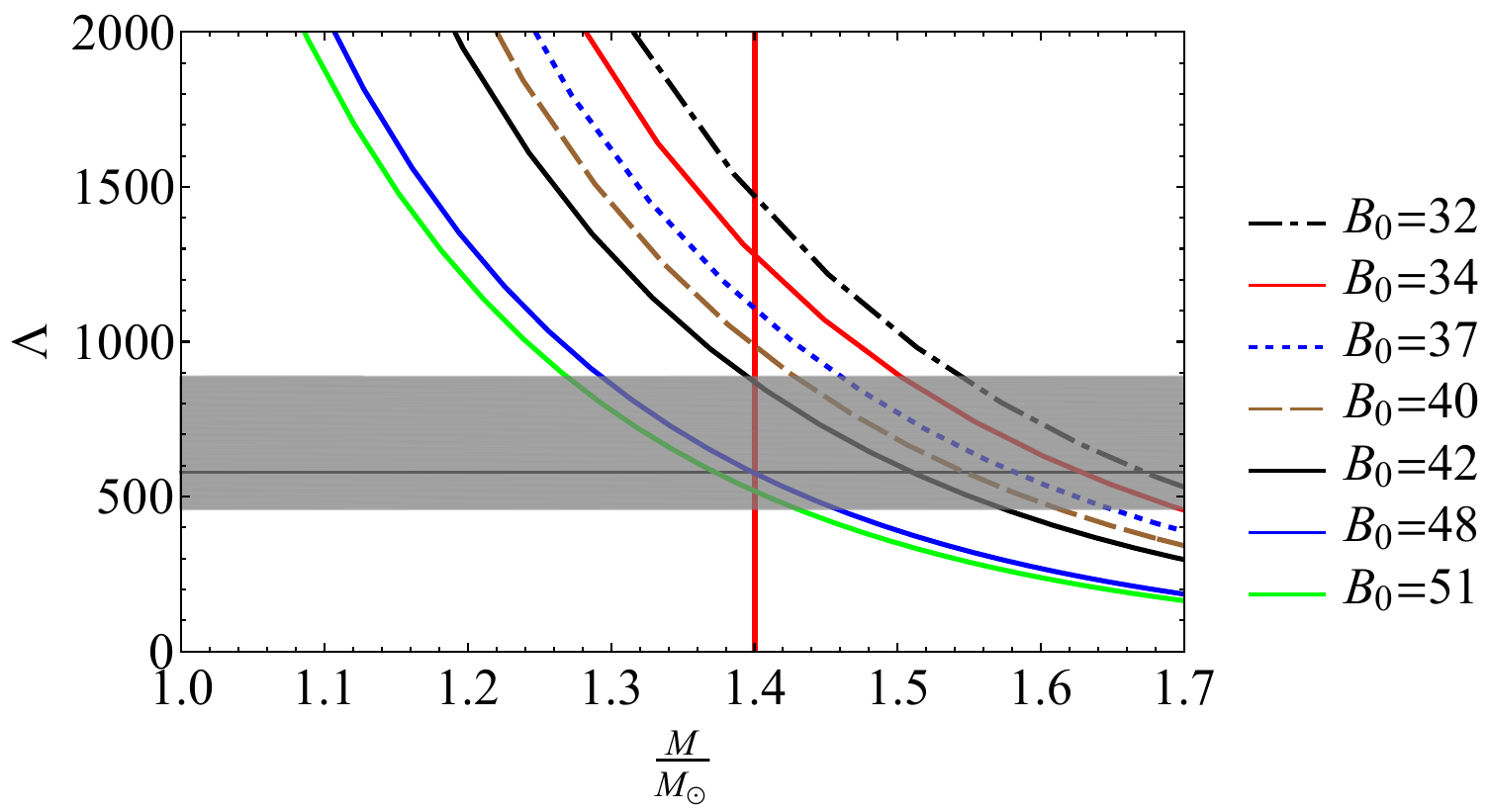}
		{a=0.4}
	\end{subfigure}
	\hfill
	\begin{subfigure}{0.45\textwidth}
		\centering
		\includegraphics[width=\textwidth]{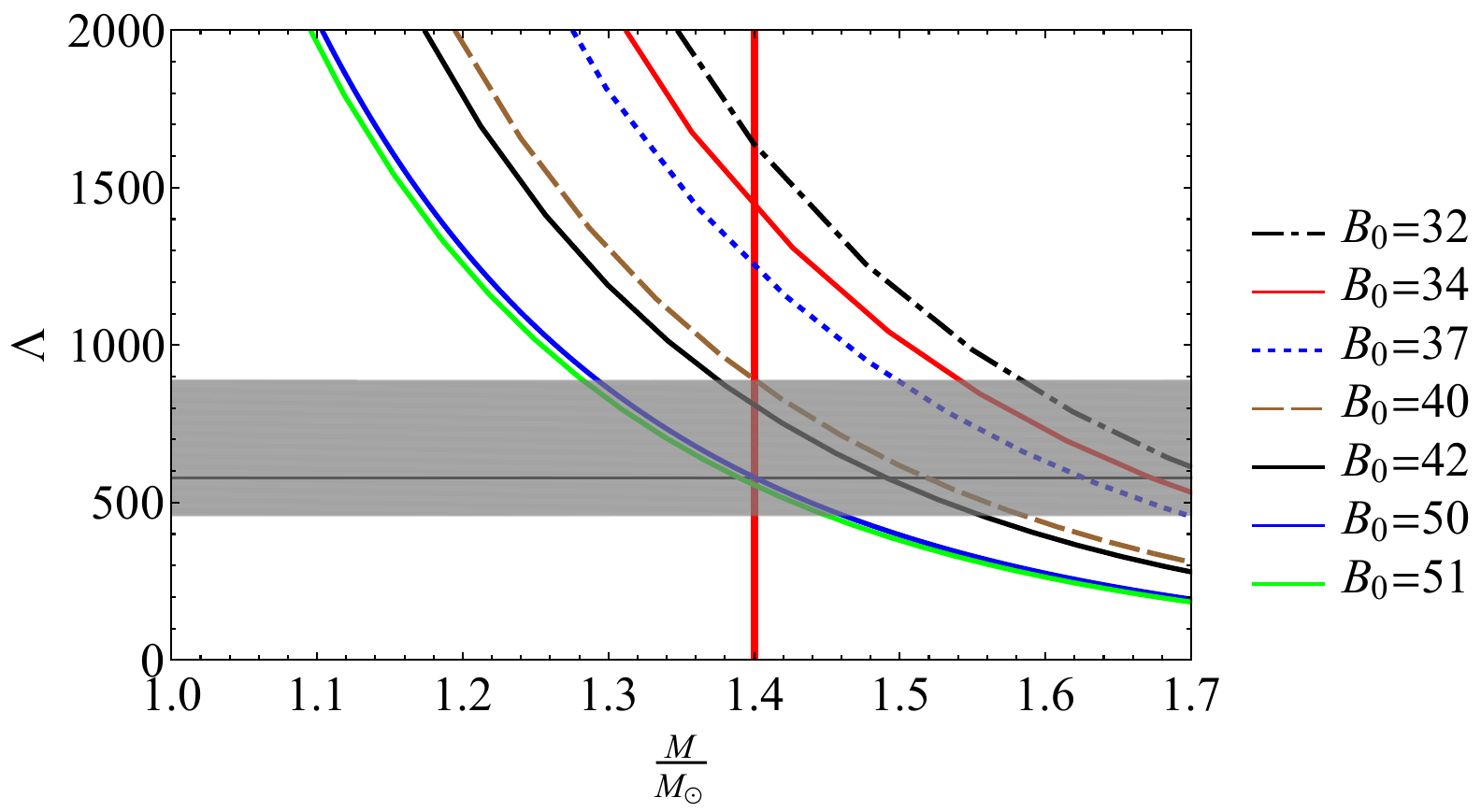}
		{a=0.8}
	\end{subfigure}
	\caption{$\Lambda-M$ diagram for different values of $a$ and $B_0$. The gray area shows the constraint $458 \lesssim \Lambda \lesssim 889$ from GW190814.}
	\label{different tidals2}
\end{figure}

To pursue further investigation, we verify the results within the bounds $458 < \Lambda_{1.4\textup{M}_\odot} < 889$ derived from GW190814. Fig. \ref{different tidals2} illustrates which results comply with this restriction. As shown in the Fig. \ref{different tidals2}, compared to Fig. \ref{different tidals}, more results fall within the gray region defined by the $\Lambda_{1.4\textup{M}_\odot}$ constraint. This is because the $\Lambda_{1.4\textup{M}_\odot}$ constraint from GW190814 includes higher values than that of from GW170817. Table \ref{results3} presents the permissible $M_{TOV}$ values corresponding to various $a$ and $B_0$ parameters, given the constraint $458 < \Lambda_{1.4\textup{M}_\odot} < 889$ derived from GW190814. As observed from the table, increasing $a$ leads to higher permissible values of $M_{TOV}$. Notably, for $a = 0.8$ and $B_0 = 40 \text{ MeV/fm}^3$, as well as $B_0 = 42 \text{ MeV/fm}^3$, the EOSs accurately capture the secondary mass of GW190814, marking these cases as particularly intriguing. Table \ref{results3} demonstrates that for these parameters, the EOSs satisfy the constraint $458 < \Lambda_{1.4\textup{M}_\odot} < 889$, and the corresponding $M_{TOV}$ falls within the mass range reported for the secondary component of GW190814 ($2.50 \textup{M}_\odot < M < 2.67 \textup{M}_\odot$).

	\begin{table}[h!]
	\caption{Structural properties of SQS for different values of $a$ and $B_0$, comparing with the $\Lambda$ constraint from GW190814.}
	\centering
	\small
	
	\begin{subtable}{1.0\linewidth}
		\centering
		\scalebox{0.77}{
			\begin{tabular}{|c|c|c|c|c|c|c|c|}
				\hline
				\multicolumn{5}{|c|}{\textbf{a=0.1}} \\
				\hline
				$B_0(MeV/fm^3)$ & $\Lambda_{1.4\textup{M}_\odot}$& $R(km)$ & $M_{TOV}(\textup{M}_\odot)$ & GW190814\\	\hline
				$32$ & 1237.10 & 13.41 & 2.53  & $\times$\\ \hline
				$34$ & 1051.20 & 12.96 & 2.45  & $\times$ \\ \hline
				$37$ & 911.72 & 12.50 & 2.36  & $\times$ \\ \hline
				$40$ & 800.32 & 12.15 & 2.30 & \checkmark \\ \hline
				$42$ & 704.62 & 11.80 & 2.23  & \checkmark \\ \hline
				\textcolor{black}{$46$}& \textcolor{black}{579.69} & \textcolor{black}{11.33} & \textcolor{black}{2.16} & \checkmark \\ \hline
				$51$ & 477.13 & 10.88 & 2.06  &  \checkmark \\ \hline
			\end{tabular}
		}
	
	\end{subtable}%
\\
	\begin{subtable}{1.0\linewidth}
		\centering
		\scalebox{0.77}{
			\begin{tabular}{|c|c|c|c|c|c|c|}
				\hline
				\multicolumn{5}{|c|}{\textbf{a=0.2}} \\
				\hline
				$B_0(MeV/fm^3)$ & $\Lambda_{1.4\textup{M}_\odot}$& $R(km)$ & $M_{TOV}(\textup{M}_\odot)$  & GW190814\\	\hline
				$32$ & 1240.59 & 13.68 & 2.63 & $\times$ \\ \hline
				$34$ & 1079.20 & 13.21 & 2.54 &$\times$ \\ \hline
				$37$ & 921.98 & 12.74 & 2.46 & $\times$ \\ \hline
				$40$ & 825.85 & 12.39 & 2.39 &\checkmark\\ \hline
				$42$ & 718.19 & 12.03 & 2.32 & \checkmark\\ \hline
				\textcolor{black}{$47$}& \textcolor{black}{576.28} & \textcolor{black}{11.44} & \textcolor{black}{2.21} &\checkmark\\ \hline
				$51$ & 494.84 & 11.08 & 2.14  &\checkmark\\ \hline
		\end{tabular}}
		
	\end{subtable}
	\\
	\begin{subtable}{1.0\linewidth}
		\centering
		\scalebox{0.77}{
			\begin{tabular}{|c|c|c|c|c|c|c|}
				\hline
				\multicolumn{5}{|c|}{\textbf{a=0.4}} \\
				\hline
				$B_0(MeV/fm^3)$ & $\Lambda_{1.4\textup{M}_\odot}$& $R(km)$ & $M_{TOV}(\textup{M}_\odot)$  & GW190814\\	\hline
				$32$ & 1465.02 & 14.28 & 2.72  & $\times$ \\ \hline
				$34$ & 1273.30 & 13.79 & 2.63 & $\times$\\ \hline
				$37$ & 1102.30 & 13.33 & 2.54  & $\times$\\ \hline
				$40$ & 986.10 & 12.94 & 2.48  & $\times$\\ \hline
				$42$ & 867.65 & 12.56 & 2.41 & \checkmark\\ \hline
				\textcolor{black}{$48$}& \textcolor{black}{577.23} & \textcolor{black}{11.68} & \textcolor{black}{2.30} &  \checkmark\\ \hline
				$51$ & 516.45 & 11.42 & 2.25  & \checkmark\\ \hline
		\end{tabular}}
		
	\end{subtable}
\\
	\begin{subtable}{1.0\linewidth}
		\centering
		\scalebox{0.77}{
			\begin{tabular}{|c|c|c|c|c|c|c|}
				\hline
				\multicolumn{5}{|c|}{\textbf{a=0.8}} \\
				\hline
				$B_0(MeV/fm^3)$ & $\Lambda_{1.4\textup{M}_\odot}$& $R(km)$ & $M_{TOV}(\textup{M}_\odot)$ & GW190814\\	\hline
				$32$ & 1644.21 & 15.01 & 2.93 & $\times$ \\ \hline
				$34$ & 1441.68 & 14.60 & 2.83  &$\times$\\ \hline
				$37$ & 1246.71 & 13.98 & 2.74  &$\times$ \\ \hline
				$40$ & 888.34 & 13.61 & 2.66  & \checkmark\\ \hline
				$42$ & 809.02 & 13.21 & 2.59  & \checkmark\\ \hline
				\textcolor{black}{$50$}& \textcolor{black}{577.39} & \textcolor{black}{12.15} & \textcolor{black}{2.43}  & \checkmark\\ \hline
				$51$ & 555.84 & 11.86 & 2.38  & \checkmark \\ \hline
		\end{tabular}}
		
	\end{subtable}
	
	\label{results3}
\end{table}	

{\section{Polytropic Models of EOS vs Perturbative QCD EOS}
In this section, we review the steps to obtain our perturbative EOS, and highlight the differences between our EOS and that of the MIT bag model. Then, we compare our EOS with the EOS of the  authors who use a generalized polytropic form with adjustable parameters. 
Our method for deriving the EOS of SQM is based on perturbative QCD, rather than the MIT Bag model. In the MIT Bag model, quark interactions are simplified as a bag constant, $B$. However, using a perturbative EOS for quark matter that includes a running coupling constant ($\alpha_s$) and a varying strange quark mass  is more accurate than that of the simple Bag model. In a purely perturbative calculation, the bag constant, $B$,  is typically set to zero. However, in more realistic models,  $B$  should not be zero, as pressure can only be defined up to an additive constant that reflects the pressure difference between the physical and perturbative vacuums  \cite{Kurkela2010, Sedaghat2021, JSedaghat}. For this reason, we treat the bag constant as a free parameter in our model, allowing us to include non-perturbative effects that are not accounted for in  weak coupling expansion. In fact, if QCD interactions between quarks are ignored, the EOS reverts to the MIT Bag model, with B acting as the bag constant \cite{Kurkela2010, Sedaghat2021, JSedaghat}. To calculate the EOS, we made the following physical assumptions:\\
i) charge neutrality; the system must remain electrically neutral locally.\\
ii) beta equilibrium; the chemical potentials are balanced by weak interactions.\\\
iii) pressure conditions; pressure must always be non-negative and should reach zero at the star's surface.\\
iv) stability of SQM; the minimum energy per baryon must be less than that of the most stable nuclei to ensure stability.\\
v) causality; the speed of sound should not exceed the speed of light.\\
vi) dynamical stability; the adiabatic index must be greater than 4/3.\\
vii) realistic quark matter; the baryon number density should exceed the saturation density.\\
viii) pressure-energy density relation; pressure and energy density should satisfy the condition $P > 0$ and $ \epsilon > P$.\\
Using these conditions and QCD calculations, we derived the EOS and compared them with observational data from LIGO and VIRGO, focusing on parameters such as mass, radius, and tidal deformability. However, some researchers have taken a different approach to determine the EOS.  They use polytropic forms of EOS with adjustable parameters.  These parameters are tuned so that the resulting EOSs meet the observational constraints. Comparing both methods could provide valuable insights. In certain studies  \cite{Maurya2023,Naeem2021,Azam2016,Azam2017}, a generalized polytropic EOS has been used, which depends on three constants $c_1$, $c_2$, and $c_3$, and a polytropic index $n$ as follows,
\begin{equation}
P=c_1 \ \epsilon^{1+1/n}+c_2 \ \epsilon + c_3
\end{equation}
By setting $c_1=0$, $c_2=1/3$, and $c_3=-4/3$, this EOS becomes that of the MIT Bag model \cite{Maurya2023}. 
Based on the eight conditions described earlier, we consider three EOSs: the MIT Bag model, the perturbative EOS with constant $B$, and the perturbative EOS with density-dependent $B$. Assuming $B=51 \frac{MeV}{fm^3}$ in both MIT bag model and perturbative EOS with constant $B$, and using $B_0=51 \frac{MeV}{fm^3}$ with $a=0.4$ for perturbative EOS with density dependent $B$, the fitting these EOSs gives the following polynomials,
\begin{equation} \label{polytrop1}
P = - 
2.48\times 10^{-9} \epsilon^3 + 8.50\times 10^{-6} \epsilon^2  + 0.32\epsilon  -68.80
\end{equation} 
\noindent \hspace{4em} (\text{MIT Bag Model})
\begin{equation} \label{polytrop2}
P = - 12.85\times 10^{-9} \epsilon^3  + 30\times 10^{-6} \epsilon^2  + 0.29\epsilon -68.68
\end{equation}
\noindent \hspace{4em} (\text{Perturbative EOS with constant $B$})
\begin{equation} \label{polytrop3}
P =  - 5.11\times 10^{-7} \epsilon^3	+ 7.76\times 10^{-4} \epsilon^2 + 0.08\epsilon -51.95.
\end{equation} 
\noindent \hspace{4em} (\text{Perturbative EOS with varying $B$})\\ \\
As demonstrated by Eqs. (\ref{polytrop1}), (\ref{polytrop2}) and (\ref{polytrop3}), we have derived the optimal fit for our EOSs using a generalized polytropic form as follows,
\begin{equation} \label{poly-eos} 
P=c_1 \ \epsilon^{3}+c_2 \ \epsilon^2 + c_3 \ \epsilon + c_4
\end{equation}
Next, we demonstrate the importance of each term in our EOSs, both with a constant and a density-dependent $B$. If we apply Eq. (\ref{poly-eos}) for the MIT Bag model, only the last two terms ($c_3 \ \epsilon + c_4$) significantly contribute to the EOS. However, when we apply the perturbative calculations and consider a density-dependent $B$, additional terms become important.
\begin{figure}[h!]
	\centering
	\begin{subfigure}{0.50\textwidth}
		\centering
		\includegraphics[width=\textwidth]{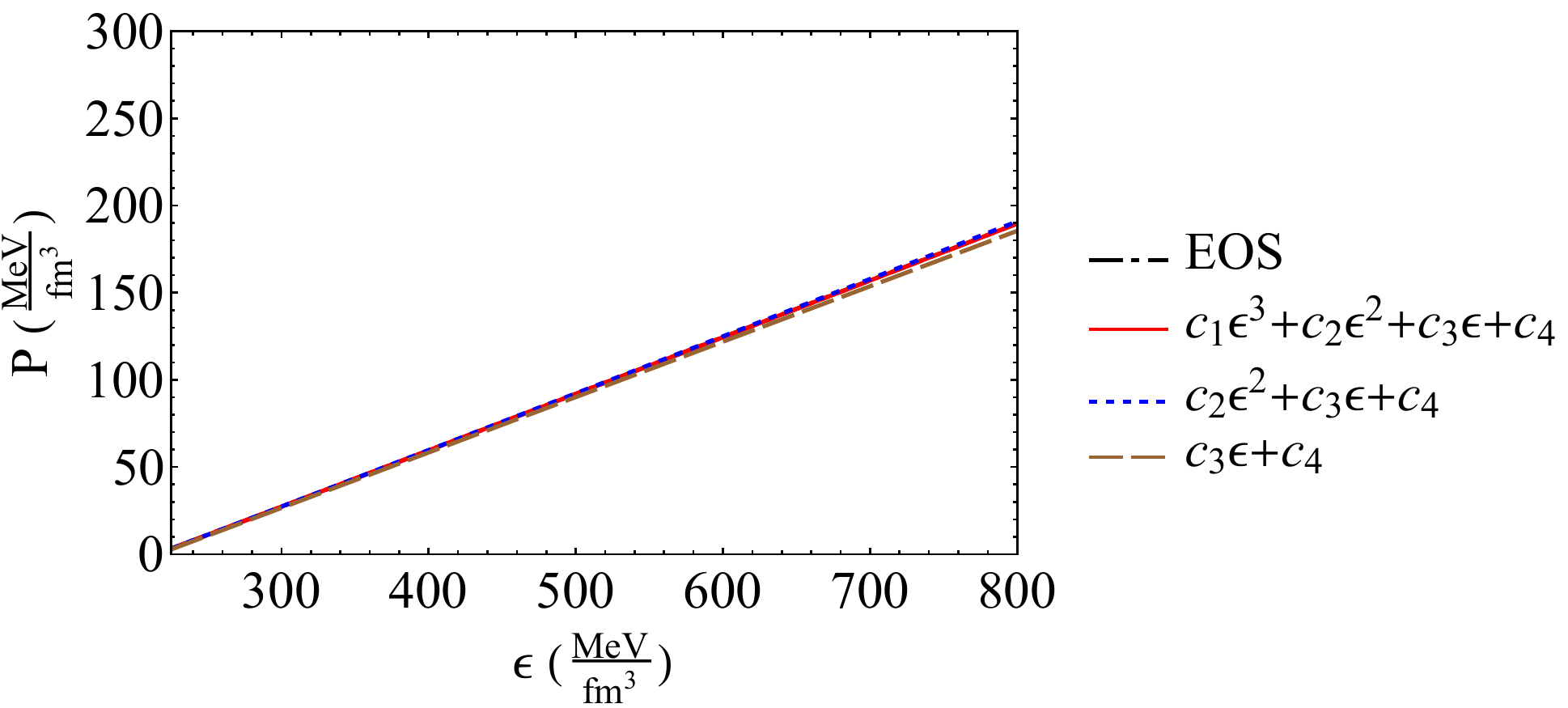}
		{a}
	\end{subfigure}
	\hfill
	\begin{subfigure}{0.50\textwidth}
		\centering
		\includegraphics[width=\textwidth]{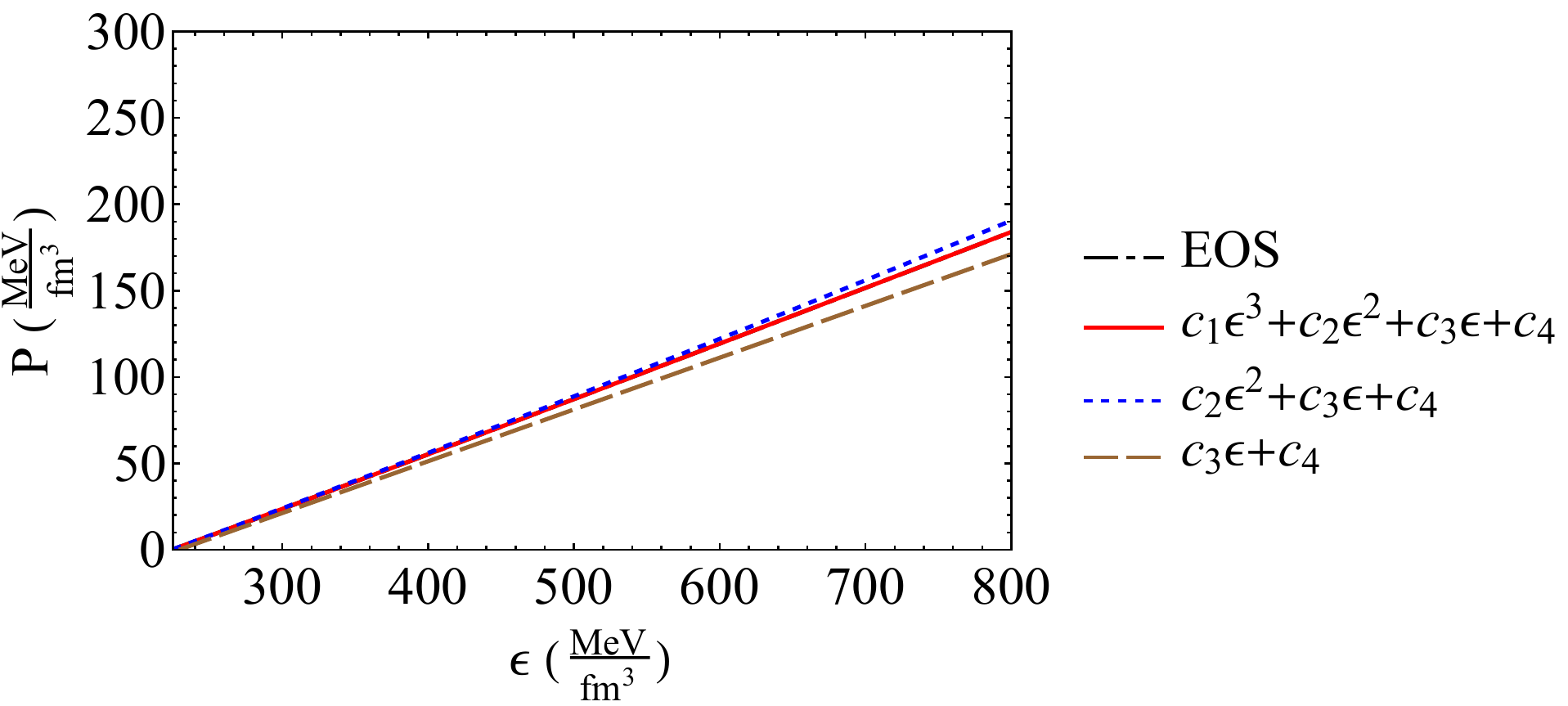}
		{b}
	\end{subfigure}
	\hfill
	\begin{subfigure}{0.50\textwidth}
		\centering
		\includegraphics[width=\textwidth]{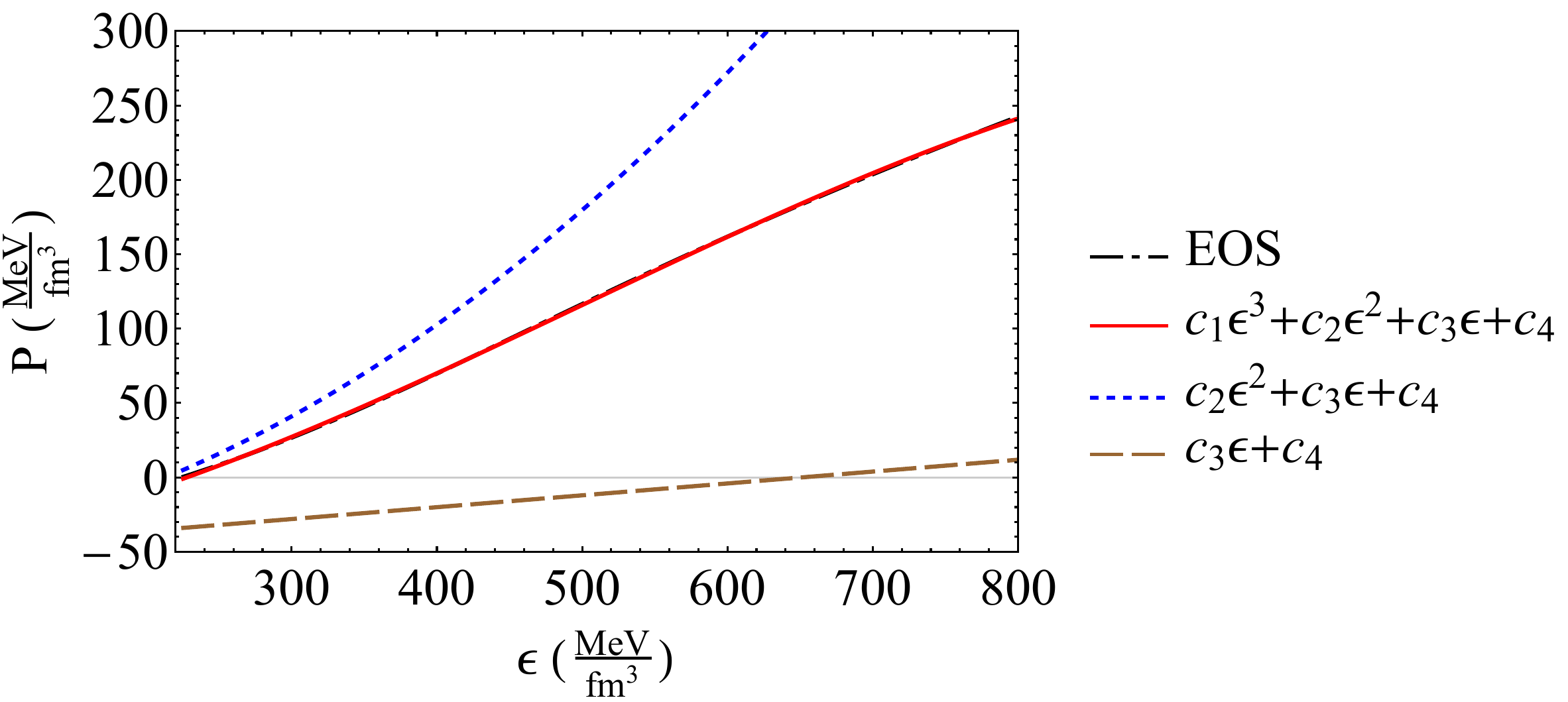}
		{c}
	\end{subfigure}
	\caption{EOSs with various contributing terms. a: MIT Bag Model, b: Perturbative EOS with constant $B$, and c: Perturbative EOS with density-dependent $B$.}
	\label{EOS123}
\end{figure}
Figure \ref{EOS123} displays these EOSs, highlighting the varying contributions of the terms in the generalized polytropic (\ref{poly-eos}). As shown in Fig. \ref{EOS123}, there is no noticeable difference between the contributing terms of the EOS derived from the MIT Bag model. However, for perturbative QCD with constant B, it is clear the terms involving up to the second power of $\epsilon$ with terms involving up to the third power of $\epsilon$  become significant. For this EOS, the these two contributions are nearly identical, suggesting that the generalized polytropic of the form $c_2 \ \epsilon^2 + c_3 \ \epsilon + c_4$ provides a good approximation for this type of EOS. Interestingly, in ref \cite{Maurya2023}, it has been shown that such a quadratic EOS can accurately describe many compact stars. The more intriguing EOS is the third one; perturbative QCD with density-dependent $B$. As shown in Fig. \ref{EOS123},  $c_2 \ \epsilon^2 + c_3 \ \epsilon + c_4$ and $c_1 \ \epsilon^3+c_2 \ \epsilon^2 + c_3 \ \epsilon + c_4$ contributions are distinctly separated, which indicates that for this type of EOS, a generalized polytropic EOS of the form $c_1 \ \epsilon^3+c_2 \ \epsilon^2 + c_3 \ \epsilon + c_4$ is the most appropriate choice. Our main contribution in this work is the development of a generalized polytropic EOS derived from QCD calculations with a density-dependent \( B \), incorporating all mentioned specified physical conditions, not by adjusting the coefficients appearing in (\ref{poly-eos}). Additionally, as discussed in previous sections, the EOSs encompass both low and high mass ranges while also satisfying tidal deformability constraints.}

\section{Discussion and conclusion}
In this paper, we have conducted a comprehensive study on the SQS using a QCD perturbative model in conjunction with the most recent dataset from the Particle Data Group. Considering the energy scale in compact stars, QCD perturbation theory on its own might be insufficient to fully describe their structure. To include non-perturbative effects, we introduced an effective bag parameter, $B$, and formulate the EOS for SQM. We initially treated $B$ as a constant parameter and later defined it as a  density-dependent function, to explore its impact on the structural properties of SQS. Furthermore, We examined the running coupling constant $\alpha_s$ and the running strange quark mass $m_s(Q)$ using the latest Particle Data Group dataset. The thermodynamic potential of SQM was derived, accounting for both non-interacting and perturbative segments, enabling us to calculate various thermodynamic properties such as pressure, energy density, quark number density, speed of sound, and adiabatic index. We first considered a constant $B$ and determined its permissible range based on the stability condition of SQM, ensuring that the energy per baryon at zero pressure is lower than that of the most stable nuclear matter. We then calculated the EOS for different values of $B$, revealing that an increase in $B$ softens the EOS, influencing the maximum gravitational mass and tidal deformability of SQS. The speed of sound and adiabatic index were computed to ensure they meet the causality and dynamical stability conditions, respectively. By solving TOV equations, we derived the mass-radius relationship and tidal deformability of SQS. Our results indicated the limitations of EOSs with a constant $B$ in describing massive objects with $M> 2.03M_{\odot}$. To achieve a more realistic depiction of SQS, we introduced a density-dependent function for $B$. This function, defined by two parameters including $B_0$ and $a$. $B_0$ is the value of bag at the surface of the SQS that is determined by stability condition of SQM. For each $B_0$, the allowed values of $a$ are determined by the constraint on tidal deformability. Our findings revealed that  incorporating  a density-dependent $B$ into the perturbative EOS can lead to SQSs with masses over $ 2M_{\odot} $, while also meeting gravitational wave requirements such as tidal deformability, and meeting to thermodynamic criteria including stability conditions and  speed of sound limits. This finding is significant as it suggests that SQS could potentially explain observations of massive compact objects, such as PSR J0952-0607, PSR J2215+5135, PSR J0740+6620, and the secondary mass of GW190814. 

Finally, it is also important to note that some studies have also utilized a density-dependent bag constant to calculate the EOS of SQM. 
A model for a density-dependent $B$ includes a Gaussian function, as detailed in Refs. \cite{Podder2024,G.F Burgio2002,Prasad2004,Bordbar2012,Pal2023}, is as follows,
\begin{eqnarray}
B(n_B) = B_\infty + (B_{n_B=0} - B_\infty) \exp \left[ -\beta \left( \frac{n_B}{n_{sat}} \right)^2 \right] \:. \label{e:g}
\end{eqnarray}
Here, $n_{sat}$ represents the saturation number density, $B_{n_B=0}$ denotes the bag pressure at zero baryon number density, $B_\infty$ is the asymptotic value of $B$, and $\beta$ is a parameter that controls the rate at which $B$ decreases with increasing $n_B$. These parameters are typically set using experimental data from CERN, specifically from studies on the formation of a quark-gluon plasma in heavy ion collisions \cite{G.F Burgio2002, Bordbar2012}. However, the medium created in these collisions differs from that in compact stars. The former occurs at low baryon density and high temperature, whereas the stable SQM is expected to form at high baryon density and low temperature. In addition, the parameters $B_{n_B=0}$ and $ B_\infty $ vary widely. For instance, in \cite{G.F Burgio2002}, $ B_{n_B=0} $ ranges from $200 \, \text{MeV/fm}^3$ to $400 \, \text{MeV/fm}^3$. In contrast in \cite{Podder2024}  $B_{n_B=0}$ varies from $50 \, \text{MeV/fm}^3$ to $ 100 \, \text{MeV/fm}^3 $ for quark stars and from $ 260 \, \text{MeV/fm}^3 $ to $ 860 \, \text{MeV/fm}^3 $ for hybrid stars. In \cite{Pal2023}, for stable SQM, the allowed range for $ B_{n_B=0} $ is $ 57 \, \text{MeV/fm}^3 $ to $ 200 \, \text{MeV/fm}^3 $, and $ B_\infty $ varies from $ 25 \, \text{MeV/fm}^3 $ to $ 50 \, \text{MeV/fm}^3 $. Additionally, according to various sources \cite{G.F Burgio2002, Prasad2004, Bordbar2012, Podder2024}, $ B_\infty $ shows a broad range, from $ 10 \, \text{MeV/fm}^3 $ to $50 \, \text{MeV/fm}^3 $. The advantages of our model compared to previous models are as follows:
I) We defined a density-dependent $B$ considering the medium in SQSs.
II) Our model for $B$ contains just two parameters, which are set by theoretical and observational constraints of compact stars.
III) The allowed range of parameters is considerably narrower than in previous models.
IV) We added $B$ as a non-perturbative contribution to the QCD perturbative EOS, allowing the strange quark mass and coupling constant to vary with energy, whereas previous models considered them as constant parameters.
V) The behavior of the sound speed in our model is well-behaved, converging to that of a non-interacting relativistic gas at ultra-high densities.
VI) Our model for the EOS of SQS matches the observational data for the mass-radius ($M-R$) relation and the tidal deformability-mass ($\Lambda-M$) relation well.
Future research should focus on refining the QCD perturbative model and exploring other non-perturbative effects that may influence the properties of SQM. Additionally, obtaining direct observational evidence for the existence of SQS remains a critical challenge.  
{It is important to highlight that we can also explore other corrections in the EOS and gravitational effects that may contribute to an increase in the star's mass. For example, accounting for the color-flavor-locked (CFL) phase can elevate both the mass and stability of quark matter \cite{Tangphati2022}. Additionally, introducing a net charge within the star can lead to a considerable increase in mass \cite{Pretel2022a,Pretel2022b}. In the context of massive gravity \cite{sedaghat} and F(R,T) gravity \cite{Tangphati2022}, the quark stars can achieve masses that are equal to or exceed the secondary mass observed in GW190814. Future studies may focus on further corrections to the EOS and the application of modified gravitational theories. In summary, by investigating the quark stars, we can improve our understanding of the EOS for dense matter which has a key role for predicting the structure, stability, and observational signatures of compact objects. In particular, these studies contribute to understanding fundamental questions about the behavior of matter under extreme conditions, including whether SQM is the true ground state of matter. Additionally, observing or constraining the properties of quark stars through gravitational wave signals, pulsar timing, and other astrophysical methods offers a window into high-energy astrophysics and may help confirm or rule out theories of exotic stars. Reviewing  references \cite{Annala, JSedaghat, Zhang2022, Banerjee2021, Zhiqiang Miao, I.Bombaci2021} reinforces the value of studying compact stars, particularly quark stars, by highlighting their unique characteristics and the fundamental insights they offer into extreme states of matter. Here, it can be said that the EOS that we have derived in this paper can be of interest, because it covers both low and high observed masses well, in addition to exhibiting appropriate sound speed behavior.}

\section*{Acknowledgements}
We wish to thank Shiraz University Research Council. This work is based upon research funded by Iran National Science Foundation (INSF) under project No. 4022870.
{ 
	\section{Appendix}
\appendix
\section{Comparative Analysis of the Bag Constant in Perturbative and MIT Bag Models}\label{appendixA}
A question that may arise for readers is why the allowed range of $B$  values, which we have used as the non-perturbative part, are smaller than the  allowed range  used in the bag model. In this appendix, we address this question in detail. Our approach for deriving the EOS of SQM relies on perturbative QCD (see sections \ref{mass and coupling} and \ref{e-p}). In this model, we treat $B$ as a free parameter to account for non-perturbative effects not captured by a simple weak coupling expansion. Thus, in the perturbative EOS, some interactions are calculated perturbatively, while the non-perturbative effects are incorporated into the effective value of the bag constant, $B$. On the other hand, the MIT Bag model simplifies all quark interactions into a single bag constant. In Section \ref{sc}, we outlined the process for deriving the permissible values of the $B$ constant in relation to the stability conditions of SQM. We established that the allowable range for $B$ is determined by ensuring that the energy per baryon at zero pressure corresponding to the star's surface remains lower than that of the most stable form of nuclear matter. We apply a similar approach for the MIT bag model. Since the mass of the strange quark is considered constant in the bag model, we assume its mass to be $100 MeV$. Figure \ref{SCBag} illustrates the range of allowable bag constant values that meet the stability criteria. It is evident that the permissible range of $B$ in the bag model is significantly wider compared to the perturbative model (see Fig. \ref{stabilitycondition}). According to perturbative calculations, the maximum allowable value for $B$ is $51 \frac{MeV}{fm^3}$, but in the simple MIT bag model, assuming a strange quark mass of $100\ MeV$, this value can increase to $83 \frac{MeV}{fm^3}$. This difference arises because, in the perturbative model, a portion of the quark interactions is derived from the perturbative expansion of the thermodynamic potential, whereas in the bag model, all the interactions are entirely characterized by the $B$ parameter.
\begin{figure}[h!]
	\centering
	\includegraphics[width=9cm,height=6cm]{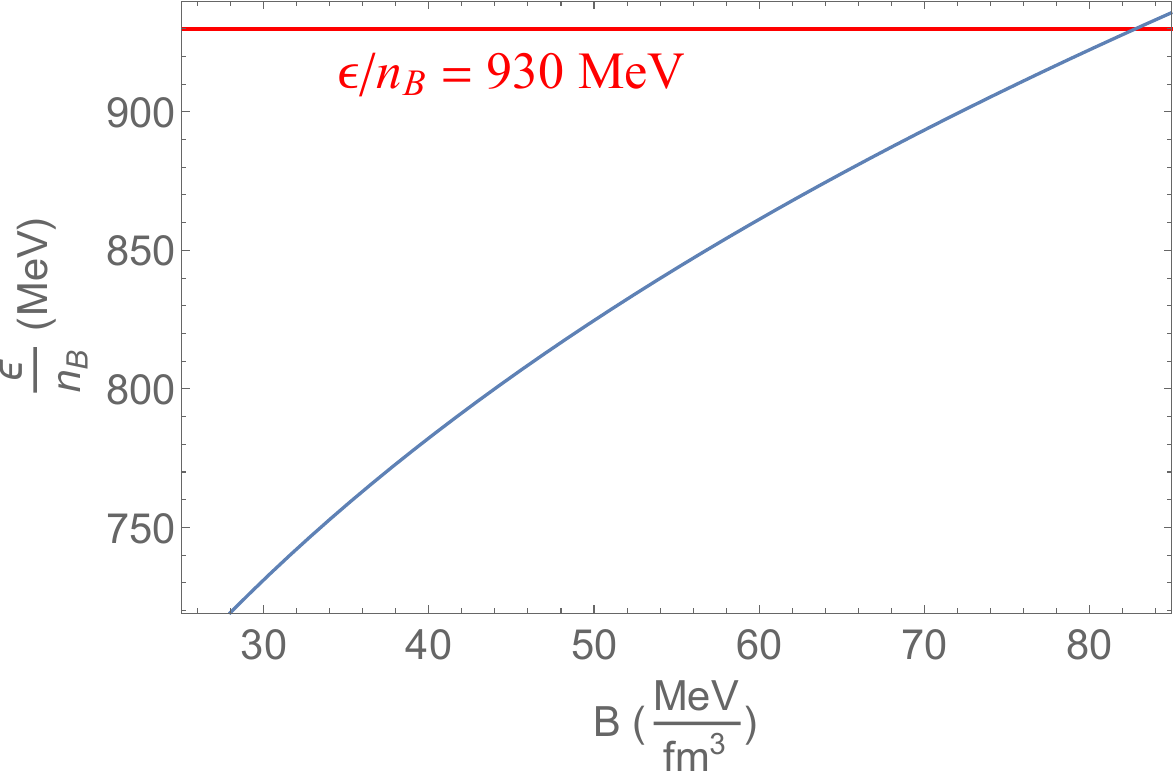}
	\caption{Stability condition in MIT bag model.}	\label{SCBag}
\end{figure}}

\end{document}